# Transonic aeroelasticity: a new perspective from the fluid mode

Chuanqiang Gao, Weiwei Zhang[*]

*School of aeronautics, Northwestern Polytechnical University, Xi'an 710072, China*

**Abstract:** Within the transonic regime, the aeroelastic problems exhibit many unique characteristics compared with subsonic and supersonic cases. Although a lot of research has been carried out in this field, the underlying mechanisms of these complex phenomena are not clearly understood yet, resulting in a challenge in the design and use of modern aircraft. This review summarizes the recent investigations on nonclassical transonic aeroelastic problems, including transonic buzz, reduction of transonic buffet onset, transonic buffeting response and frequency lock-in phenomenon in transonic buffet flow. After introducing the research methods in unsteady aerodynamics and aeroelastic problems, the dynamical characteristics as well as the physical mechanisms of these phenomena are discussed from the perspective of the fluid mode. In the framework of the ROM (reduced order model) -based model, the dominant fluid mode (or the eigenvalue) and its coupling process with the structural model can be clearly captured. The flow nonlinearity was believed to be the cause of the complexity of transonic aeroelasticity. In fact, this review indicates that the complexity lies in the decrease of the flow stability in the transonic regime. In this condition, the fluid mode becomes a principal part of the coupling process, which results in the instability of the fluid mode itself or the structural mode, and thus, it is the root cause of different transonic aeroelastic phenomena.

**Keywords:** aeroelasticity, transonic buffet, transonic buzz, lock-in, fluid mode, fluid-structure interaction

# 1. Introduction

The design of modern aircraft, especially military aircraft, generally pursues the requirements of high speed, high maneuverability but light weight. This will lead the aircraft to be at the potential risk of various transonic aeroelastic problems. In addition to the classical bending-torsion flutter, transonic aeroelasticity also exhibits many unique phenomena [1-3]. For example, the sweptback wing experiences a drop of the flutter boundary in the transonic regime, called the transonic dip phenomenon [4-7]. There is a relatively large discrepancy between the computational and experimental flutter boundaries when the Mach numbers are near the transonic dip [8]. The control surface (rudder or aileron) may encounter self-excited, and often limit-cycle oscillations in the single degree of freedom during flight at transonic or low supersonic speeds, which is the transonic control surface buzz phenomenon [9-12]. In exceptional cases, the wing may even display a kind of nodal-shaped oscillation caused by the interaction between flutter and buffet in transonic flow [13-16]. In transonic buffet flow conditions, furthermore, the frequency of the structural response may not follow the buffet frequency but locks onto the natural frequency of the structure, causing the frequency lock-in phenomenon, in which a large oscillating amplitude of the wing can be observed in the lock-in region [17-20].

---

[*] Corresponding author E-mail: aeroelastic@nwpu.edu.cn

Because of the abundant and complex phenomena, transonic aeroelasticity has received continuous attentions from academia and industry. In this period, wind tunnel experimentation is one of the main approaches to conduct transonic aeroelasticity research. Many valuable conclusions were obtained by this method. Even now, it is still an important means in designing a new airplane and studying complex transonic aeroelastic problems. Besides, some transonic aeroelastic experiments are used as standard cases for the validation of the numerical simulation, of the transonic flutter experiment with the AGARD445.6 wing [4], the control surface buzz experiment with the NASP wing [21], the transonic static aeroelastic experiment with the HIRENASD wing [22], and the benchmark active control technology (BACT) experiment with NACA0012 airfoil section [23]. Since Steger [24] carried out the study of transonic buzz by solving the NS equations in 1980, numerical simulation has gradually become the main approach in transonic aeroelasticity investigations. In the past 40 years, researchers proposed many approaches to deal with the problems in CFD/CSD time domain simulation, i.e. the coupling scheme, data exchange method on the fluid-solid interface and dynamic grid technology. Although the CFD/CSD simulation method has achieved great progress, it is still inefficient and difficult to obtain the response, and to reveal the physical mechanism of complex transonic aeroelastic phenomena. In order to address these drawbacks, some researchers have attempted to construct a reduced order model (ROM) analysis based on modal coordinates to replace the CFD/CSD simulation since 2000 [25-26]. In recent years, the ROM-based method has received wide attention. A variety of modeling methods, including linear and non-linear, have been proposed in the fields of unsteady aerodynamics and aeroelasticity.

In the classical concept of aeroelasticity, dynamic aeroelastic problems are usually classified into two types: aeroelastic stability problem and dynamic response problem. The first one, commonly referred to as flutter, is the most critical issue in the field of aeroelasticity. It is of great significance in engineering. Typical representatives include classical bending-torsion flutter and transonic buzz. The bending-torsion flutter is the nature of coupling instability between structural modes. The flutter boundary usually displays a dip in the transonic region, which makes the transonic flutter boundary a limitation of the flight envelope. The transonic buzz refers to the limit-cycle oscillations of the control surface around its hinge. It is a kind of special flutter, only the single structural degree of freedom involved. For aeroelastic stability problems, the coupling methods, considering the bidirectional coupling feedback between flow and structure, are necessary in order to actually illustrate the dynamic characteristics of the FSI system. The freestream flow is considered stable in most researches.

Unlike the stability problem in stable flow, the dynamic response problems focus on the flow-induced vibration of the wing due to unstable or unsteady flows. In transonic buffeting the excitation source is the aerodynamic force fluctuations caused by the global instability of the shock boundary-layer interaction flow [27]. Buffeting vibration will not only affect the fatigue life of the aircraft, but also have undesirable impacts on the operation of the airborne instrumentation and of the flight control system. Because the self-sustained oscillation of the shock wave is independent of the motion of the structure, the feedback effect between fluid and structure is considered to be too weak to perform a coupling analysis from the traditional aeroelastic viewpoint. In addition, considering the limitations on strength and stiffness of the scaled model in the wind tunnel experiment, the decoupled "two-step method" is adopted as the main approach to study the transonic buffeting problem, The analysis process of the two-step method is first to predict the buffeting loads based on the rigid wing, and then calculate the response of the elastic wing under the loads. The key is to accurately predict the buffeting loads. As a result, transonic buffeting, also referred as transonic buffet, is usually regarded as a pure fluid mechanics phenomenon. A large number of wind tunnel experiments and numerical simulations on transonic buffet have been carried out with rigid wings/airfoils by many institutions and scholars [27-34].

The above classification is more dependent on the stability of the flow itself. Then the corresponding research methods are adopted according to the classification, as an aeroelastic stability problem or as a dynamic response problem. However, this research route has obvious limitations. Firstly, transonic buzz is classified as a stability problem, thus, most researchers pay little attention to the effect of the flow stability. Although some phenomena observed are difficult to explain from the viewpoint of the traditional flutter, engineers still use it to guide the design of the control surface system. Some scholars believe that the occurrence of the transonic buzz is closely related to transonic buffet, but there is little direct evidence and systematic research on this point [35]. Moreover, another limitation lies in the field of transonic buffet. It is classified as a dynamic response problem, in which the coupling effect is often neglected and in which rigid wings are used in most current investigations with the uncoupled two-step method. Within this framework, the frequency lock-in phenomenon cannot be explained from the resonance theory [20,36]. Consequently, the physical mechanisms of the non-classical transonic aeroelastic problems remain unclear. This is the root cause why modern aircraft still suffer from transonic aeroelastic problems.

In fact, the structure of the aircraft is usually elastic, and it is impossible to guarantee the absolute stability of the flow when the aircraft flies in transonic regime. Therefore, there is no clear boundary between the two classifications, and it is very difficult for researchers to distinguish whether the flow is stable or not. To deal with this issue, it is necessary to construct a unified analysis method for both kinds of problems, and further to understand the underlying mechanisms from a unified perspective.

There have been several review papers on aeroelasticity. However, they paid less attention to the mechanism of transonic aeroelastic problems. For example, Dowell [37-38] and Afonso [39] separately summarized the aeroelastic phenomena from the viewpoint of nonlinearity. Although the transonic cases were considered, the mechanisms of some phenomena were still not well explained [40]. Nonlinearity is far too general to predict the specific flow conditions and/or structural parameters for the occurrence of the non-classical transonic aeroelastic issues. This means it fails to provide a clear design criterion for engineers to avoid these problems. Transonic aeroelasticity, as a result, is still a long-term obstacle in the field of aerospace. In 2011, Bendiksen [2] summarized the research progress of transonic unsteady aerodynamics, in which transonic aeroelasticity was only one of the application cases. Therefore, there is a pressing need to summarize the recent progress on transonic aeroelasticity, especially the physical mechanism of certain non-classical transonic aeroelastic problems. Our group has carried out long-term research in this area. On the basis of the achieved progresses, this paper reviews the construction of the unified analysis method and the mechanism from the perspective of fluid mode towards typical transonic aeroelastic phenomena.

This paper is organized as follows. Section 2 provides an overview of the numerical methods for transonic unsteady flow, which includes the CFD approach and the reduced-order model (ROM). For the transonic buffet flow cases, the applicability of simulations in URANS is first discussed, and then the framework of the ROM with the ARX method as well as the modeling steps are presented. The aeroelastic analysis methods, including CFD/CSD time domain simulation and the ROM-based aeroelastic model are introduced in Section 3. In Section 4, the dominant fluid mode, representing the stability and frequency spectrum of the flow, obtained by ROM is introduced. From the perspective of fluid mode, this review summarizes the dynamical characteristics and physical mechanisms of certain non-classical transonic aeroelastic problems, including transonic buzz, reduction of transonic buffet onset, classical transonic buffeting response and the frequency lock-in phenomenon in transonic buffet flow. Conclusions are provided in Section 5.

## 2. Methods for the transonic unsteady flow

The complexity of transonic aeroelasticity has its origin in the complex transonic flow which can be divided into three levels, namely, steady flow, unsteady flow, and the unstable flow caused by the global instability [3]. Transonic steady flow, dominated by the spatial nonlinearity, refers to the nonlinear characteristics of the transonic steady flow varying with flow parameters, i.e. freestream angle of attack. In this case the objective is to accurately predict the strength of the shock wave. Transonic unsteady flow is caused by the structural motion. If the structure vibrates with a large amplitude, the motion of the shock wave (and the aerodynamic load) are no longer linear with the structural motion. The linear relationship is achieved only in the case of the structural motion with a sufficiently small amplitude. Under the assumption of small disturbance, this kind of unsteady flow can be converted to a time-linearized problem, which is the foundation of many aeroelastic studies. The third level is caused by the global instability. The shock wave can oscillate even in the absence of the structural motion [41]. In this kind of unstable flow, the interaction between the structural motion and the self-excited shock oscillation can result in complex aeroelastic phenomena when an elastic wing is adopted. Nonlinearity, however, makes it very difficult to understand the underlying mechanisms of these aeroelastic phenomena. A ROM-based aeroelastic analysis method is an important way to deal with such problem.

## 2.1 Numerical simulation by CFD

CFD is one of the most important means for flow analysis. Thanks to the contribution of several generations of fluid mechanics scientists, numerical solutions of the N-S equations have achieved great progress in the transonic flow simulation. Unsteady Reynolds-averaged Navier-Stokes (URANS), Detached-Eddy Simulation (DES) and Large Eddy Simulation (LES) have been successively applied to the simulation of transonic buffet flow.

Although the LES method can more precisely obtain the evolution process of separated vortices, it is not widely used in the transonic buffet simulation due to its inefficiency. Deck [42] used the standard DES and Zonal-DES methods to study transonic buffet flow over the OAT15A airfoil, and compared them with URANS. He found that Zonal-DES could obtain more accurate buffet characteristics, while the standard DES and URANS obtained comparable results on buffet onset and buffet loads. Later, Chen [43], Huang [44], Grossi [45-46] and Sartor [47] also applied the hybrid RANS/LES method to calculate the transonic buffet, proposing improved approaches such as DDES and IDDES. In general, the improved DES methods yield more accurate unsteady flow characteristics and separated regions than URANS methods. However, most of the improved methods need to set zones with different grid topologies in advance according to the flow features, and then to identify the RANS or LES method in each zone. When the shock location and the separation region change due to the structural motion, the zones have to be redefined to get a reasonable result. Because of this limitation, the hybrid RANS/LES method is not convenient for the study of the fluid-structure interaction problem.

The URANS method was proved to be particularly sensitive to the choice of turbulence model to calculate the transonic buffet flow. For the NACA0012 airfoil, Barakos [48] studied the effectiveness of various linear and nonlinear eddy viscosity models, including the Baldwin−Lomax model, the Spalart−Allmaras (S−A) model, linear and nonlinear $k$ - $\varepsilon$ models and a nonlinear $k$ - $\omega$ model. The author found that only the S-A model and $k$ - $\omega$ model developed unsteady flow characteristics comparable to the experiment, but at a slightly larger angle of attack. Goncalves [49] conducted a similar investigation based on the supercritical RA16SC1 airfoil. He also found the S-A model could reproduce the buffet unsteadiness with reasonable buffet frequencies but the amplitude were underpredicted. In addition to the turbulence model, a number of authors have reported similar sensitivity studies on the numerical discretization scheme, the time steps and grid resolution [50-54]. These studies have accumulated experience for the URANS simulation of transonic buffet. It is well known that the buffeting characteristics of airfoils are mainly dominated by the large-amplitude motion of the shock wave and the resulting large-scale separated vortices. Furthermore, periodic shock motions in transonic buffet occur in time scales that are much larger than those of the wall-bounded turbulence [55-56]. Therefore, with an appropriate turbulence model, the URANS method is suitable for the simulation of transonic unsteady problems, i.e. transonic buffet flow.

Using the URANS method and the S-A turbulence model, Gao et al. [57] calculated the transonic buffet onset on the NACA0012 airfoil, in which the inviscid flux was discretized by the second-order advection upstream splitting method (AUSM) scheme, and the viscous flux term is discretized by the standard central scheme. It is known that buffet onset is a combination of Mach number and angle of attack, which represents the boundary of the shock changing from static to periodic oscillation. Figure 1 shows the comparison of the buffet onset boundary, in which the circles represent the data from the experiment conducted by Doerffer et al. [30] and the solid line represents the computational results from the present URANS method. As can be seen, the experimental data and URANS calculations are in good agreement. At the Mach number of 0.70, the calculated buffet onset angle is 4.80 degrees, which is very close to the 4.74 degrees obtained from experiment. Figure 2 shows reduced buffet frequencies at different angles of attack at $M = 0.7$. As the angle of attack increases, the buffet frequency increases smoothly and slightly. At $\alpha = 5$ degrees, the reduced frequency is 0.180 (about 0.058 at the scale of Strouhal number, $S_t = f_b c / U_\infty$), which is very close to the experimental data of 0.176. The reduced frequency is defined as $k_b = \pi f_b c / U_\infty$, in which $f_b$ is the buffet frequency; $c$ indicates the chord of the airfoil, and $U_\infty$ denotes the velocity of the free-stream.

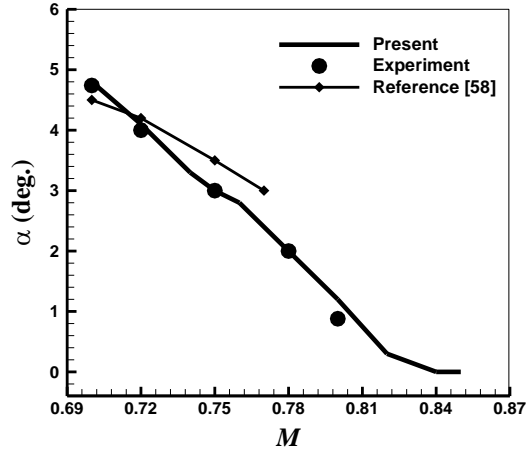

Figure 1. Comparison of the buffet onset boundary between URANS simulation and experiment

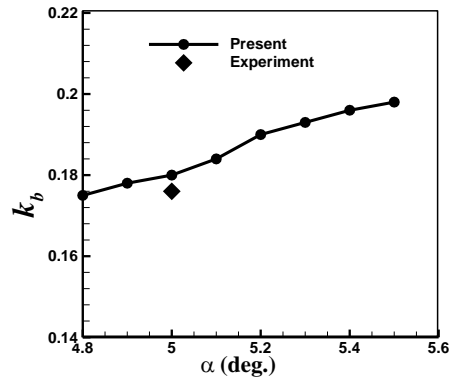

Figure 2. Buffet frequencies at different angles of attack [57]

Gao et al. [57] also verified the transonic buffet loads based on a supercritical OAT15A airfoil. This model has been applied to conduct a series of wind tunnel experiments in the state of $M = 0.73$, $\alpha = 3.5$ degrees by Jacquin [31]. Figure 3 shows the root mean square (RMS) values of the pressure fluctuation on the upper profile surface, where $Q_0$ is the free-stream dynamic pressure. In the present calculation, the predicted peak at 3.5 degrees is slightly lower than that of the experiment, whereas the calculated peak at 3.7 degrees matches better. However, the result simulated by Deck [42] using Zonal-DES at 3.5 ° shows a higher peak. This is consistent with the experience shared by other researchers. That is, the buffet load predicted by the URANS method is slightly smaller. The angle of attack, therefore, is usually slightly increased in URANS simulation in order to match the experiment data. Fig. 4 displays the velocity contours and the velocity profiles at different phases in the location of $x/c$=0.6, where $U_0$ is the free-stream velocity and $y_s$ is the coordinate of the upper surface. Phase 1 is defined at the moment when the shock is in the upstream location. The boundary layer flow is attached at phase 5 (Fig. 4(b)), and the separation occurs at phase 9 (Fig. 4(c)). In the given four phases, velocity profiles obtained by calculations agree well with those of experiment. From the above validations, the URANS method with S-A turbulence model can provide a high-accuracy prediction of the transonic buffet onset and load.

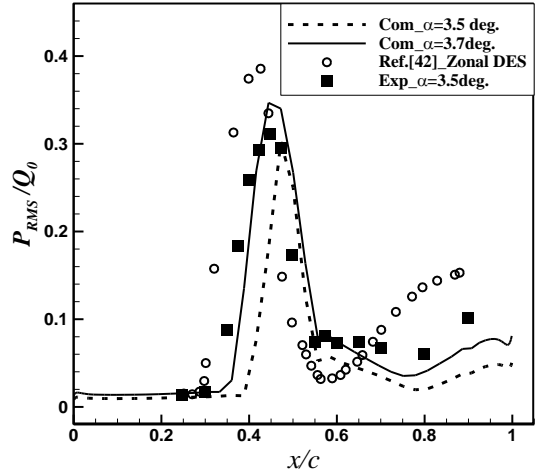

Figure 3. Comparison of the RMS of the pressure between simulation and experiment [57]

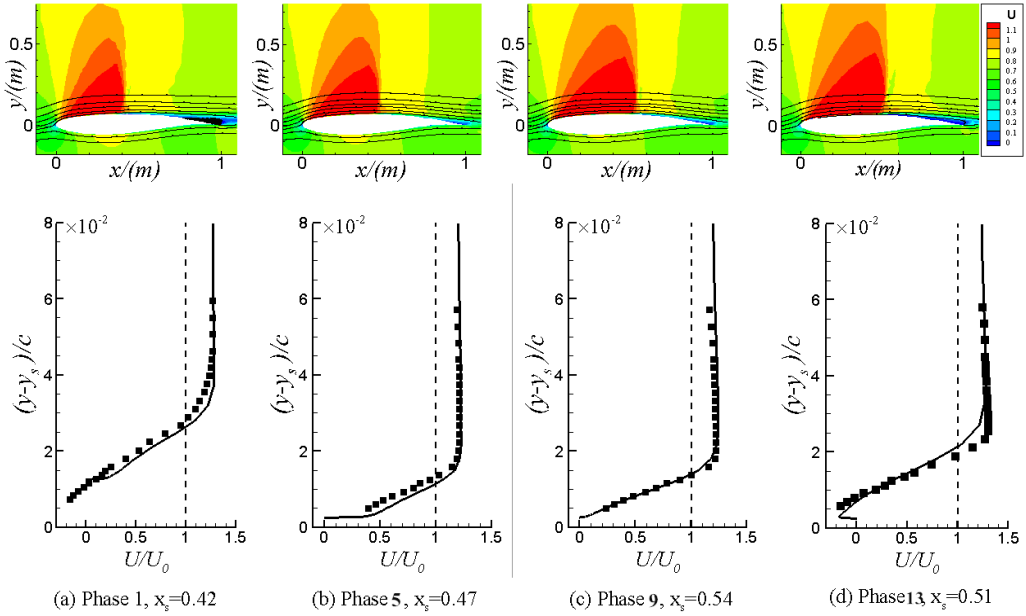

(a) Phase 1, $x_s$=0.42   (b) Phase 5, $x_s$=0.47   (c) Phase 9, $x_s$=0.54   (d) Phase13, $x_s$=0.51

Figure 4. The velocity contours and velocity profiles of four typical phases in one period [57]

## 2.4 Reduced-order model for the unsteady flow

CFD simulation can accurately predict the flow features of the unsteady flow and its evolution. Nevertheless, this technique is often inefficient, because it must solve the complex N-S equations. It becomes even more inefficient when applied to the fluid-structure interaction (FSI) problem, in which a coupling of two sets of equations is involved. ROM of the unsteady flow is an alternative approach to overcome the above shortcomings of CFD. The nature of ROM is to construct a low-order model that can reproduce the main dynamic characteristics of the flow to replace the full-order CFD approach [26,59]. Since the 1990s, researchers have developed a variety of ROMs for the unsteady flow based on CFD data, and some have been successfully applied in the fields of aeroelasticity, flight dynamics, aerodynamic optimization design and flow control. According to the modeling framework, the current

ROMs can be divided into two categories: the modal decomposition method based on the flow feature extraction and the identification method based on input-output samples.

By projecting the high-dimensional training samples onto several characteristic bases, the flow dynamics can be represented and reconstructed with a reduced dimensionality, which is the core idea of the modal decomposition method. This methodology mainly contains proper orthogonal decomposition (POD), dynamic mode decomposition (DMD), harmonic balance (HB) and Jacobian-based Taylor expansion techniques. POD determines the low-order subspaces through an orthogonal transformation process. After combination with the flow governing equations [60-62] or surrogate models [63-65], it is easy to approximate and analyze the flow dynamics with less time cost and a high level of accuracy. The DMD method was recently proposed based on the infinite-dimensional linear operator description. This method is able to decompose the flow into several modes with a single frequency and growth rate, which has been extensively used to study linear, periodic and even nonlinear flow cases [66-69]. The harmonic balance method allows a direct calculation of the periodic state, which has been used for both aeroelastic simulations [70] and dynamic derivative predictions [71]. The Jacobian-based Taylor expansion methods also belong to projection-based methods as the full-order system is projected onto a basis formed by a small number of eigenvectors of the Jacobian matrix, which have been successfully used in the control design of flexible aircraft [72].

System identification-based methods aim to provide linear or nonlinear algorithms that reveal the underlying relationship of input-output data. These approaches describe a flow system with a brief mathematical expression, thus decreasing the computational cost. Compared with the projection-based methods, they are easy to establish because only a few input-output samples, rather than the full order flow snapshots, are needed. Therefore, these approaches are well suited for dealing with problems of dynamic responses like flutter. For linear modeling processes, Volterra series, Eigensystem Realization Algorithms (ERA) and AutoRegressive with eXogenous input (ARX) are commonly used identification methods. Silva [73] developed ROM for the unsteady aerodynamics using Volterra series, and accomplished the aeroelastic stability analysis of the AGARD445.6 wing. Yao and Jaiman [74] used ERA to analyze the vortex-induced vibration of blunt bodies. Zhang applied the ARX method to study flutter [75-76], frequency lock-in in vortex-induced vibrations [77], transonic buffet [20] and aeroelastic optimization [78-79].

As to the problems of nonlinear aeroelasticity, i.e. limit cycle oscillation under the nonlinearity of a large-amplitude structural motion or flow separation, practical nonlinear ROMs have been developed, such as the Volterra series model [80], the neural network model [81], and the Winer model [82]. To improve the generalization ability of these nonlinear ROMs and construct linear/non-linear combination models, Kou [83-84] and Winter [85-87] have carried out extensive original researches. In recent years, increasing attention has been paid to deep learning and machine learning methods in this field [88].

Although great progress has been achieved in the nonlinear reduced-order methodologies of unsteady aerodynamics, the linear ROM is still a very important analytical method for aeroelastic problems due to its incomparable advantages in revealing the physics mechanism. Therefore, a linear ROM constructed by the ARX method is adopted in the following review, in which the input is the generalized displacement of the structure and the output-1 is the corresponding generalized aerodynamic force.

For a general multiple input and multiple output (MIMO) system, the ARX approach can be expressed in the form of discrete difference equations as follows:

$$y(k) = \sum_{i=1}^{na} A_i y(k-i) + \sum_{i=1}^{nb-1} B_i u(k-i) + e(k) \tag{1}$$

where $y(k)$ means the system output vector at $k^{\text{th}}$ step and $u(k)$ is the system input at the same step. $A_i$ and $B_i$ are the constant coefficients to be estimated by the least-squares method. To ensure that the mean value is zero, constant levels need to be first removed from the sampled data. $na$ and $nb$ are the orders of the model chosen by the user. Both parameters are dependent on the unsteady characteristics of the flow to be identified. For the flow with a weaker unsteady effect, such as the supersonic flow, a second-order model ($na=nb=2$) is sufficient enough to accurately predict the unsteady aerodynamic loads. In general, the stronger the unsteady effect is, the larger the order that is required.

In order to construct the coupled model, the difference equation (1) is transformed into the state space form. The state vector $x_a(k)$ consisting of ($na+nb-1$) states is defined as follows:

$$x_a(k) = [f_a(k-1), \cdots, f_a(k-na), u(k-1), \cdots, u(k-nb+1)]^T \tag{2}$$

The discrete-time aerodynamic model in state-space form is as follows:

$$\left. \begin{array}{l} x_a(k+1) = \tilde{A}_a x_a(k) + \tilde{B}_a u(k) \\ f_a(k) = \tilde{C}_a x_a(k) + \tilde{D}_a u(k) \end{array} \right\}, \tag{3}$$

where:

$$\tilde{A}_a = \begin{bmatrix} A_1 & A_2 & \cdots & A_{na-1} & A_{na} & B_1 & B_2 & \cdots & B_{nb-2} & B_{nb-1} \\ I & 0 & \cdots & 0 & 0 & 0 & 0 & \cdots & 0 & 0 \\ \vdots & I & \cdots & 0 & 0 & 0 & 0 & \cdots & 0 & 0 \\ \vdots & \vdots & & \vdots & \vdots & \vdots & \vdots & & \vdots & \vdots \\ 0 & 0 & \cdots & I & 0 & 0 & 0 & \cdots & 0 & 0 \\ 0 & 0 & \cdots & 0 & 0 & 0 & 0 & \cdots & 0 & 0 \\ 0 & 0 & \cdots & 0 & 0 & I & 0 & \cdots & 0 & 0 \\ 0 & 0 & \cdots & 0 & 0 & 0 & I & \cdots & 0 & 0 \\ \vdots & \vdots & & \vdots & \vdots & \vdots & \vdots & & \vdots & \vdots \\ 0 & 0 & \cdots & 0 & 0 & 0 & 0 & \cdots & I & 0 \end{bmatrix}$$

$$\tilde{B}_a = \begin{bmatrix} \tilde{B}_0 & 0 & \cdots & 0 & I & 0 & 0 & \cdots & 0 \end{bmatrix}^T$$

$$\tilde{\boldsymbol{C}}_a = \begin{bmatrix} \boldsymbol{A}_1 & \boldsymbol{A}_2 & \cdots & \boldsymbol{A}_{na-1} & \boldsymbol{A}_{na} & \boldsymbol{B}_1 & \boldsymbol{B}_2 & \cdots & \boldsymbol{B}_{nb-2} & \boldsymbol{B}_{nb-1} \end{bmatrix}$$

$$\tilde{\boldsymbol{D}}_a = \begin{bmatrix} \boldsymbol{B}_0 \end{bmatrix}$$

Then the discrete-time equation (3) is turned into the continuous-time form equation (4) by the bilinear transformation. With this model, the characteristics of the flow can be obtained by analyzing of the eigenvalues of matrix $\boldsymbol{A}_a$.

$$\left. \begin{aligned} \dot{\boldsymbol{x}}_a(t) &= \boldsymbol{A}_a \boldsymbol{x}_a(t) + \boldsymbol{B}_a \boldsymbol{u}(t) \\ \boldsymbol{y}(t) &= \boldsymbol{C}_a \boldsymbol{x}_a(t) + \boldsymbol{D}_a \boldsymbol{u}(t) \end{aligned} \right\} \tag{4}$$

The above ARX method has been widely used to construct the ROM in the aeroelastic analysis. It contains two steps: sample training and parameter identification. However, most studies focus on the modeling of the stable flow under a small disturbance satisfying the linear assumption. The well-developed transonic buffet flow is unstable and nonlinear. The growing amplitudes of the aerodynamic forces will ultimately give rise to a limit cycle behavior. Under this condition, the training process will exacerbate the instability and nonlinearity of the system. Although a nonlinear modeling method can be adopted to reproduce the buffeting response, we prefer to construct a linear model and then to analyze the underlying mechanism of these complex aeroelastic problems. Previous studies have shown that the ARX method is suitable to construct ROMs for the unstable flow as long as the sample training process meets the linear assumption, namely, the following two requirements:

1) The sample training must be calculated based on the steady-state base flow. That is, the base flows in pre-buffet and buffet conditions need to be obtained in advance.

2) The training signal should be designed with an appropriate bandwidth and amplitude so as to avoid the nonlinear behavior in the training.

We set up the aerodynamic model at $M=0.7$, $\alpha=5.5$ degrees to verify the ROM. In order to accurately predict the unsteady aerodynamics, delay orders should be assigned with relatively large numbers. The error is defined as follows:

$$e = \frac{\sum_{i=1}^{L} |\boldsymbol{y}(i) - \boldsymbol{y}_{iden}(i)|}{\sum_{i=1}^{L} |\boldsymbol{y}(i)|}, \tag{5}$$

where $\boldsymbol{y}_{iden}$ represents the vector of the identified aerodynamic forces, and $L$ is the length of the training signals. The identified results with $na=nb=80$ are compared with those of CFD simulations in Figure 5; and a good agreement is observed between them.

The ROM is then validated in the time domain by comparing with CFD simulations under the excitation of *harmonic signals*. Figure 6 shows the comparison of time histories of the pitching moment coefficient between ROM and the CFD simulation at two typical forcing frequencies ($n=0.7$ and $n=1.3$). ROM predictions exactly match those computed by the CFD simulation, especially when the amplitude of the pitching moment coefficient is less than 0.015.

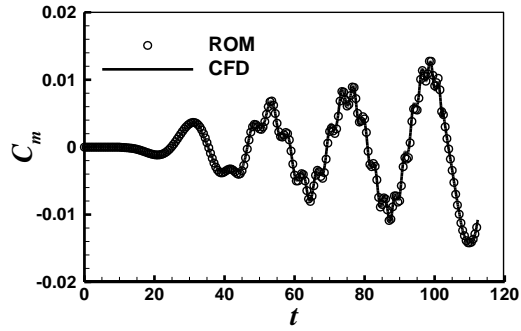

Figure 5. Identified results compared with those of CFD simulations. [20]

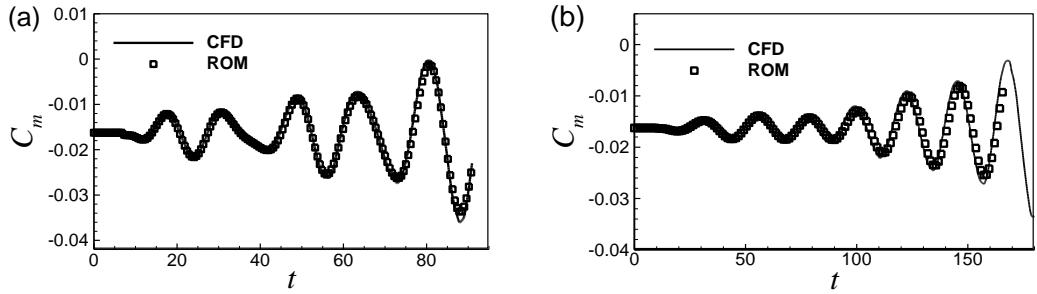

Figure 6. Comparison of time responses of a harmonic pitching motion at α=5.5degrees for (*a*) $\theta = 0.02\sin(1.4\pi f_b t)$ and (*b*) $\theta = 0.02\sin(2.6\pi f_b t)$ . [20]

# 3. Numerical methods for aeroelastic problems

Similar to the aerodynamic simulation, the aeroelastic analysis also adopts two approaches, namely, the CFD/CSD time domain simulation and the ROM-based aeroelastic model. Their relationship is shown in Figure 7. The first approach is the CFD/CSD time domain simulation (S-F1), in which the aerodynamic response is obtained by the CFD simulation (URANS method), and the structural motion equation is solved in the time domain. The other one is the ROM-based aeroelastic model, in which a ROM is constructed to obtain the generalized aerodynamic loads and the aeroelastic equation is established in the state space.

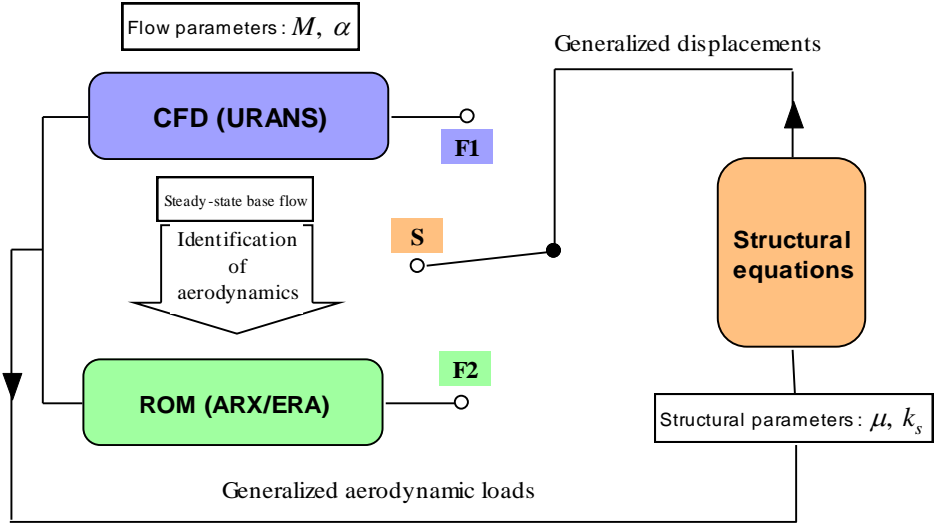

Figure 7. Aeroelastic analysis methods and their relationship

## 3.1 CFD/CSD simulation in time domain

Since Steger [24] applied the CFD method to the transonic aeroelastic simulation, the CFD/CSD time domain simulation has become one of the most important approaches in this field. Under this framework, the generalized aeroelastic motion equation is written in matrix form as follows:

$$\boldsymbol{M}\ddot{\boldsymbol{\xi}} + \boldsymbol{G}\dot{\boldsymbol{\xi}} + \boldsymbol{K}\boldsymbol{\xi} = \boldsymbol{Q} \tag{6}$$

where $\boldsymbol{\xi}$ are the generalized displacement, generalized velocity and generalized acceleration, respectively; $\boldsymbol{M}$ represents the generalized mass. $\boldsymbol{K}$ is the generalized stiffness. $\boldsymbol{G}$ is the generalized damping (often $\boldsymbol{G} = \boldsymbol{0}$ in flutter analysis), and $\boldsymbol{Q}$ is the generalized aerodynamic force calculated by the CFD simulation.

By defining the structural state-vector $\boldsymbol{x} = [\xi_1\ \xi_2 \cdots \xi_N\ \dot{\xi}_1\ \dot{\xi}_2 \cdots \dot{\xi}_N]^T$, the structural motion equation in state-space form is as follows:

$$\dot{\boldsymbol{x}} = \boldsymbol{f}(\boldsymbol{x},t) = \boldsymbol{A}_s\boldsymbol{x} + \boldsymbol{B}_s\boldsymbol{Q}(\boldsymbol{x},t), \tag{7}$$

where $\boldsymbol{A} = \begin{bmatrix} \boldsymbol{0} & \boldsymbol{I} \\ -\boldsymbol{M}^{-1}\boldsymbol{K} & -\boldsymbol{M}^{-1}\boldsymbol{G} \end{bmatrix}$; $\boldsymbol{B} = \begin{bmatrix} \boldsymbol{0} \\ \boldsymbol{M}^{-1} \end{bmatrix}$; $\boldsymbol{0}$ is the zero square matrix; and $\boldsymbol{I}$ is the identity matrix. Eq. (6) is a first-order differential equation system.

The fourth-order accuracy hybrid linear multi-step method [89] is used to solve the aeroelastic Eq. (7) in the time domain as follows:

$$\begin{cases} \boldsymbol{Q}_n = \boldsymbol{Q}(\boldsymbol{x}_n, t_n) \\ \boldsymbol{F}_n = \boldsymbol{A}\boldsymbol{x}_n + \boldsymbol{B}\boldsymbol{Q}_n \\ \tilde{\boldsymbol{x}}_{n+1} = \boldsymbol{x}_n + \dfrac{\Delta t}{24}(55\boldsymbol{F}_n - 59\boldsymbol{F}_{n-1} + 37\boldsymbol{F}_{n-2} - 9\boldsymbol{F}_{n-3}) \\ \tilde{\boldsymbol{Q}}_{n+1} = 4\boldsymbol{Q}_n - 6\boldsymbol{Q}_{n-1} + 4\boldsymbol{Q}_{n-2} - \boldsymbol{Q}_{n-3} \\ \tilde{\boldsymbol{F}}_{n+1} = \boldsymbol{A}\tilde{\boldsymbol{x}}_{n+1} + \boldsymbol{B}\tilde{\boldsymbol{Q}}_{n+1} \\ \boldsymbol{x}_{n+1} = \boldsymbol{x}_n + \dfrac{\Delta t}{24}(9\tilde{\boldsymbol{F}}_{n+1} + 19\boldsymbol{F}_n - 5\boldsymbol{F}_{n-1} + \boldsymbol{F}_{n-2}) \end{cases}, \tag{8}$$

where the superscript "~" is the predictor value; the subscript "$n$" is the index of the $n$th step; and $\Delta t$ is the time step.

## 3.2 ROM-based aeroelastic model

Although the time-domain CFD/CSD simulation can obtain the detailed aeroelastic system responses, it is inefficient and inconvenient to carry out the mechanism analysis. Therefore, it is necessary to introduce a low-order aeroelastic analysis model to reveal the mechanism of complex aeroelastic phenomena.

Defining the structural state-vector $\boldsymbol{x}_s = [\theta, \dot{\theta}]^T$, the structural motion equation (6) in the state-space form can be rewritten as:

$$\begin{cases} \dot{\boldsymbol{x}}_s(t) = \boldsymbol{A}_s \boldsymbol{x}_s(t) + q\boldsymbol{B}_s y_a(t) \\ u(t) = \boldsymbol{C}_s \boldsymbol{x}_s(t) + q\boldsymbol{D}_s y_a(t) \end{cases}, \tag{9}$$

where $\boldsymbol{A}_s = \begin{bmatrix} 0 & 1 \\ -k_s{}^2 & 0 \end{bmatrix}$, $\boldsymbol{B}_s = \begin{bmatrix} 0 \\ 1 \end{bmatrix}$, $\boldsymbol{C}_s = \begin{bmatrix} 1 & 0 \end{bmatrix}$, $\boldsymbol{D}_s = \begin{bmatrix} 0 \end{bmatrix}$, $q = \dfrac{1}{\pi\mu r_\alpha^2}$.

By defining the state-vector $\boldsymbol{x}_{ae} = [\boldsymbol{x}_s, \boldsymbol{x}_a]^T$ and coupling structural state equations (9) with aerodynamic state equations (4), we obtain the state equation and output equation for the linear aeroelastic system as follows:

$$\begin{cases} \dot{\boldsymbol{x}}_{ae}(t) = \begin{bmatrix} \boldsymbol{A}_s + q \cdot \boldsymbol{B}_s \boldsymbol{D}_a \boldsymbol{C}_s & q \cdot \boldsymbol{B}_s \boldsymbol{C}_a \\ \boldsymbol{B}_a \boldsymbol{C}_s & \boldsymbol{A}_a \end{bmatrix} \cdot \boldsymbol{x}_{ae}(t) = \boldsymbol{A}_{ae} \cdot \boldsymbol{x}_{ae}(t) \\ u(t) = \begin{bmatrix} \boldsymbol{C}_s & 0 \end{bmatrix} \cdot \boldsymbol{x}_{ae}(t) \end{cases}, \tag{10}$$

The ROM-based aeroelastic model is then obtained. The problem of the aeroelastic stability is converted into the analysis of the complex eigenvalues of $\boldsymbol{A}_{ae}$. The process of the transonic aeroelastic analysis is:

1) Use the URANS flow solver to compute the aerodynamic coefficient of the mode with the designed input signal in Figure 8.

2) Use the identification technique to construct the input–output difference model [equation (3)] in the discrete-time domain, and then turn it into the continue-time state-space form [equation (4)].

3) Couple the aerodynamic state space equation (4) and the structural state space equation (9), to obtain the ROM-based aeroelastic state-space equation (10).

4) Solve the complex eigenvalues of the state matrix $\boldsymbol{A}_{ae}$ in equation (10). We focus on the two most unstable eigenvalues, which are corresponding to the fluid and structural modes. With different structural parameters, natural frequencies of the elastic airfoil, and mass ratios, we can get a set of eigenvalues. It is the root loci. The stability of the aeroelastic system changes when the root loci cross the imaginary axis.

# 4. Typical transonic aeroelastic phenomena and their underlying mechanisms

The classical flutter is usually caused by the coupling of two or more structural modes, in which the flow acts like an "adhesive". The freestream density (or the dynamic pressure) is the key parameter to the occurrence of flutter. This issue has been widely studied, and the physics mechanism behind the coupling process is relatively clear. Besides the classical flutter, there also exists a couple of non-classical flutter in the transonic flow condition, like transonic buzz. The instability characteristics of such phenomena are quite different from those of the classical flutter, that is, the Mach number and angle of attack play a major role in the flutter boundary rather than the dynamic pressure. Very recent research has been performed on these non-classical aeroelastic phenomena, in which the dominant fluid mode is first derived from the ROM and then the coupling mechanism is investigated from the perspective of the fluid mode.

## 4.1 The dominant fluid mode

Like the structural dynamics system, the fluid dynamics system can also be decomposed with several fluid modes according to the modal analysis. The fluid modes are a group of inherent patterns (including modal frequency, damping ratio and modal shape) that represent the nature of the flow [90]. The main characteristic of an unsteady flow is the superposition of these fluid modes. The above-mentioned ROMs both can produce the main information about the fluid mode. The derived dominant fluid modes are essentially the same even though they are described in different ways. In identification methods, the main characteristics of the fluid modes can be obtained by solving the eigenvalues of the matrix $A_a$. The real part of the eigenvalue represents the damping ratio of the fluid modes (the positive value means that the flow is unstable), and the imaginary part represents the modal frequency. The mode shape, however, is not directly obtained from this method but from the modal decomposition method which can also acquire the corresponding damping ratio and frequency by appropriate conversion.

For the NACA0012 airfoil, the transonic buffet case at M=0.7, a=5.5 degrees is selected to illustrate the fluid modes. Figure 8 displays the eigenvalues of matrix $A_a$ in the ARX model with different orders. It can be seen that a pair of conjugate eigenvalues always lies in the right half-plane (positive real part), and it is approximately convergent with the identified accuracy. In addition, the imaginary parts of this pair are nearly equal to 0.2 (in the scale of reduced frequency), which coincides with the buffet frequency calculated by the CFD simulation. Therefore, this pair of eigenvalues correspond to the global dominant mode of the transonic buffet flow, and the dynamics of the buffet flow system is determined by this mode.

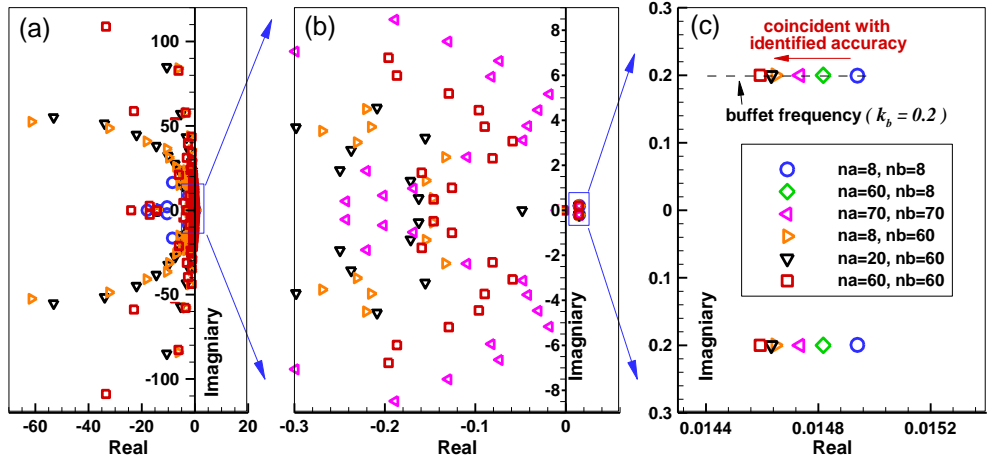

Figure 8. Eigenvalues calculated from ARX model with different orders [113]

Based on the same case, the fluid modes obtained from the DMD method are shown in Figure 9, and the corresponding reduced frequencies and growth rates are shown in Table 1. It can be noticed that both the growth rate and the frequency are zero for the first mode. It is a static mode, close to the mean flow field. All the other modes reflect the oscillating features resulting from the shock waves. The reduced frequency of Mode 2 is 0.196, which is equal to the buffet frequency from the CFD simulation. Other mode frequencies are twice or three times larger than the basic buffet frequency. Therefore, Mode 2 is the most important coherent global mode, which is the modal shape corresponding to the dominant eigenvalue by the ARX-based ROM.

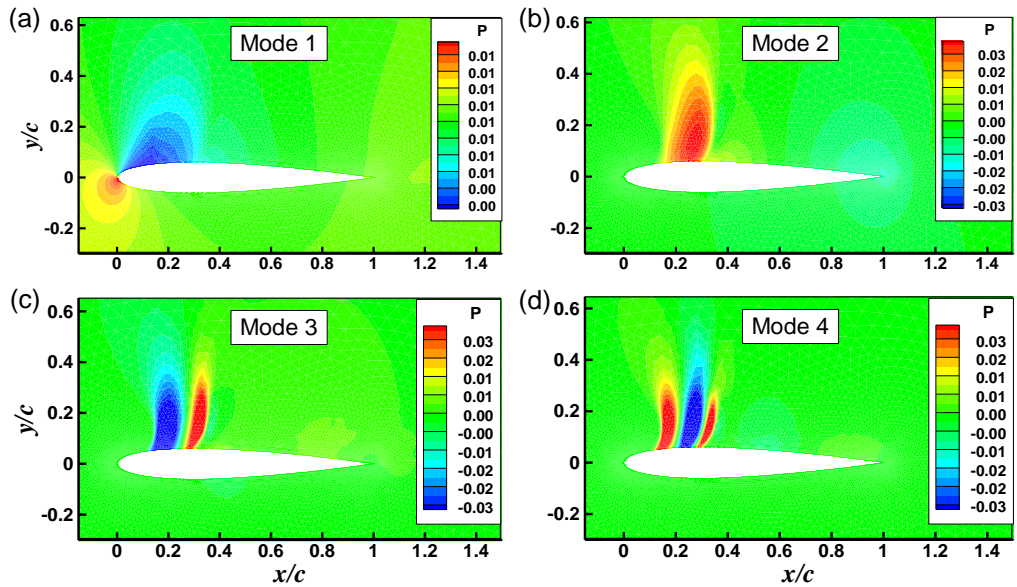

Figure 9. The first four modal shapes of the DMD mode in the buffet case of M=0.7, a=5.5 degrees [113]

Table 1 The growth rate and reduced frequency of the first four DMD modes

| Mode number | Growth rate | Reduced frequency |
|:---:|:---:|:---:|
| 1 | 0 | 0 |

| 2 | $3.75 \times 10^{-6}$ | 0.196 |
| 3 | $3.86 \times 10^{-5}$ | 0.393 |
| 4 | $1.20 \times 10^{-6}$ | 0.588 |

It is well known that the flow stability is closely related to the Mach number and the angle of attack of the freestream. Figure 10 shows the eigenvalues of the dominant global mode obtained by the ARX-based ROM at different angles of attack when the Mach number is fixed at 0.7. As the angle of attack increases, the eigenvalue of the dominant mode gradually approaches the imaginary axis, which indicates that although the flow is stable, its stability margin decreases (subcritical state). In the vicinity of 4.7 degrees, the eigenvalue moves from the left half plane to the right one. It means that the buffet onset angle is about 4.7 degrees, which is consistent with the CFD simulation and wind tunnel test. In the post-buffet conditions, the eigenvalue crosses the imaginary axis again when it is near 5.9 degrees, which corresponds to the offset angle of buffet. Then with the further increase of the angle of attack, both the real part (representing stability) and imaginary part (representing frequency) slightly increase again. Therefore, the buffet onset and offset angles of attack are 4.7 and 5.9 degrees at a Mach number of 0.7 for the NACA0012 airfoil. The strongest buffeting loads are obtained at 5.5 degrees because it has the largest real part of eigenvalue.

Furthermore, the eigenvalue of the dominant global mode is also investigated at Mach numbers from 0.80 to 0.87 at zero angle of attack. The result is shown in Figure 11. With the increase of Mach number, the modal eigenvalues also cross the imaginary axis twice. The buffeting start boundary and exit boundary are M = 0.82 and M = 0.85, respectively, which are basically consistent with the CFD simulation results.

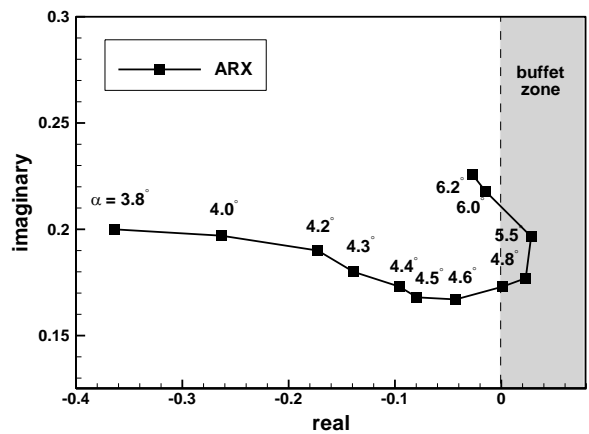

Figure 10. Eigenvalue of the dominant fluid mode predicted by ROM at different angles of attack

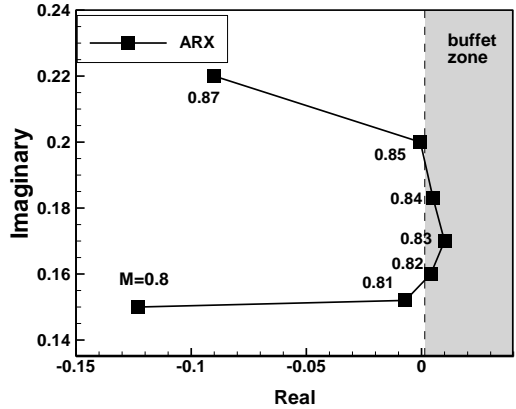

Figure 11. Eigenvalue of the dominant fluid mode predicted by ROM as a function of Mach number

The fluid modes not only provide a good way to reveal the evolution of complex unstable flows, but also open a gate to understand the coupling process of the FSI system. Kou [91] studied the vortex-induced vibration (VIV) phenomenon of a circular cylinder from the perspective of fluid mode. It is well known that the flow around a stationary cylinder becomes unstable at the Reynolds number of 47, accompanied by the periodic vortex shedding phenomenon. However, when the cylinder is elastically supported, the periodic vortex shedding can occur at a subcritical Reynolds number as low as 18. Kou explained why the lowest Reynolds number for the VIV to occur is 18 through a ROM-based FSI analysis. By performing DMD on the flow systems, the fluid mode and its corresponding eigenvalues were captured. A pair of clear and dominant fluid mode arises from Reynolds number 18 and it becomes unstable at 47. In this region, the VIV phenomenon is triggered by the FSI between this dominant fluid mode and the structure. When the Reynolds number is lower than 18, it fails to capture a definite fluid mode, thus the fluid system is nearly impossible to interact with the elastic structure, and the VIV will not happen. Therefore, the fluid mode is a new perspective to investigate the physical mechanism of complex FSI problems.

## 4.2 Coupling patterns between fluid mode and structural mode

From the perspective of dominant fluid modes, it can be seen that there exist relatively large subcritical regions near the transonic buffet onset and offset boundaries. In these conditions, although the flow is stable, the stability margin is low (the modal damping is small). It is easier to trigger the interaction between the fluid mode and the structural mode, changing the stability of the coupled system. Previous studies have found that typical aeroelastic problems, i.e. transonic buzz and frequency lock-in phenomena often occur in the vicinity of transonic buffet onset. However, these studies failed to systematically provide the intrinsic mechanisms of those phenomena and their relationships.

The transonic aeroelastic problems can be classified into four types according to the stability of the freestream flow and dynamic characteristics of the coupled FSI system, as shown in Table 2. When the freestream flow is in a subcritical state, i.e. the pre-buffet condition, the aeroelastic system will display two different instability types caused by the coupling effect. The first one is the instability in the structural mode (Type I), of which the representative case is transonic buzz, also called transonic single degree of freedom flutter. The other one is the instability in the fluid mode (Type II), which refers to the reduction of transonic buffet onset when the structural elasticity is activated. The aeroelastic phenomena in the unstable buffet flow can also be divided into two types. The first type is the instability only in the fluid mode (Type IV), namely, the forced response of the structure under the buffeting load. The other is the simultaneous instability in both the structural mode and the fluid mode (Type III), which is the frequency lock-in phenomenon in the transonic buffet flow. In the following sections, we will review the research progress of these types of non-classical transonic aeroelastic problems, and further reveal the mechanisms of these phenomena from the perspective of the fluid mode.

Table 2 Types of FSI system and corresponding FSI problems in the transonic regime

| Flow state | Pre-buffet conditions (subcritical instability) | | Buffet conditions (Instability) | |
|---|---|---|---|---|
| Types of FSI system | Type I instab. in S mode | Type II instab. in F mode | Type III retain instab. in F mode | Type IV instab. in both S & F modes |
| FSI problems | transonic buzz | reduction on transonic buffet onset | buffeting response | lock-in in transonic buffet flow |

## 4.3 Type I: Instability on structural mode in the pre-buffet flow (transonic buzz)

As early as 1947, Erickson et al. [92] conducted a series of studies on transonic buzz by means of wind tunnel experiments and theoretical analyses, and they believed that this kind of limit-cycle oscillation was a result of the generation of the shock wave and its motion on the wing surface. It was found to be caused by the phase difference between shock wave motion and aileron rotation. Lambourne [9] thought the main cause of buzz was the negative aerodynamic damping effect caused by the phase difference. These studies at the time showed that control surface buzz was related to the shock wave and the corresponding boundary layer separation, as well as the aerodynamic nonlinearity [93-94].

Bendiksen [95] discussed the range of Mach number for the potential occurrence of the control surface buzz from the relationship between the shock position and Mach number by calculating the steady flow. Furthermore, he reviewed the correlation among the flow parameters, such as, Mach number, angle of attack. These studies were conducted based on the forced vibration method, and the explanations were derived from the viewpoint of energy, that is, the transonic flow injected energy into the structure. In the review, he pointed out that although the characteristics of the single-degree-of-freedom flutter (transonic buzz) had been fully investigated, the in-depth underlying mechanism was still not clear. All aeroelastic stability problems can be explained from the perspective of energy. Nevertheless, the particularity of transonic buzz is not reflected in this explanation. Its occurrence is related not only to the flow condition, but also to the structural parameters. He et al. [96] studied the influence of the distance between the aerodynamic center and the pitching axis on the dynamic characteristics of transonic buzz. In fact, they focused on the correlation between the pitching static derivative and the occurrence of transonic buzz. It is convenient to understand the static stability of the system from the pitching static derivative. Nevertheless, transonic buzz is a dynamics problem. The instability range predicted by the pitching static derivative does not coincide with that calculated by the CFD/CSD simulation. Therefore, most of the former studies failed to reasonably explain why transonic buzz occurred in certain flow conditions and with a certain structural stiffness, which are the most important issues for the aircraft engineers.

Different from the qualitative interpretation of transonic buzz, Gao et al. [11] performed comprehensive studies on the dynamic characteristics, occurrence conditions and the underlying mechanism of this kind of non-classical flutter. The study method is the ROM-based aeroelastic model proposed in section 3.2, and the CFD/CSD simulation method is also adopted to verify the conclusions. In their study, an all-moving tail airfoil is considered, namely, the NACA0012 airfoil with a pitching spring support. The Mach number is fixed at 0.7, and the angle of attack is in the range from 4.0 degrees to 4.8 degrees. For the rigid airfoil, the buffet onset angle is 4.8 degrees at a Mach number of 0.7, and the reduced frequency of the buffeting flow is 0.17 in the onset condition. Table 3 shows the real part (damping) and the imaginary part (reduced frequency) of the dominant fluid mode predicted by the ROM in typical cases. In these conditions, the shock wave does not reach the trailing edge yet. The resulting aeroelastic problems can be classified as A/B type buzz.

Table 3 typical cases (M=0.7) and the eigenvalues of the dominant fluid mode

| α (degrees) | Real | Imaginary |
|---|---|---|
| 4.0 | -0.26 | 0.192 |
| 4.2 | -0.17 | 0.19 |
| 4.5 | -0.05 | 0.170 |
| 4.8 （Buffet onset） | 0.03 | 0.173 |

First of all, the root loci of the coupled system obtained by the ROM are shown as a function of the structural frequency in a pre-buffet case of M=0.7, a=4.5 degrees, as shown in Figure 12. The eigenvalues can be divided into two branches, the fluid mode branch and the structural mode branch that is near the imaginary axis. When the structural frequency is close to the characteristic frequency of the dominant fluid mode, "repulsion" occurs between the two branches. The eigenvalues of the structural branch cross the imaginary axis and enter into the right half plane, resulting in instability on the structural mode (flutter). The cause of the SDOF flutter (buzz), therefore, is caused by the coupling between the structural mode and the subcritical fluid mode. Figure 13 shows the "boot-shaped" instability region as a function of the mass ratio. The left boundary of the instability boot gradually gets close to the fluid characteristic frequency $k_f = 0.17$ with the increase of the mass ratio; while the right boundary almost remains around 0.42 despite changes of mass ratios. Figure 13 also presents flutter boundaries from the CFD/CSD simulation at certain mass ratios, which perfectly match the ROM results.

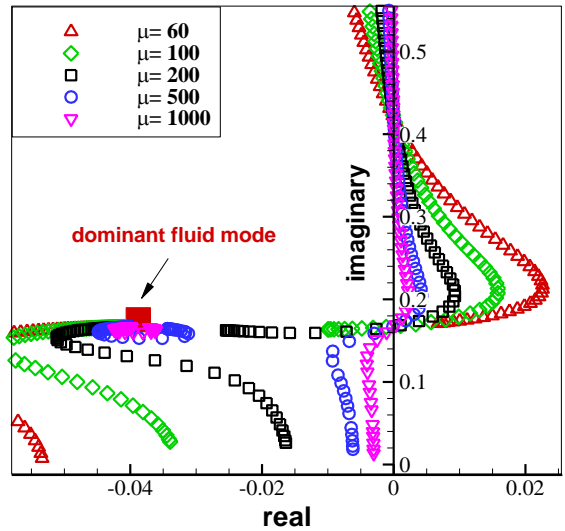

Figure 12. The root loci of the aeroelastic system with a spring pitching support at M=0.7 and $\alpha$=4.5 degrees [11]

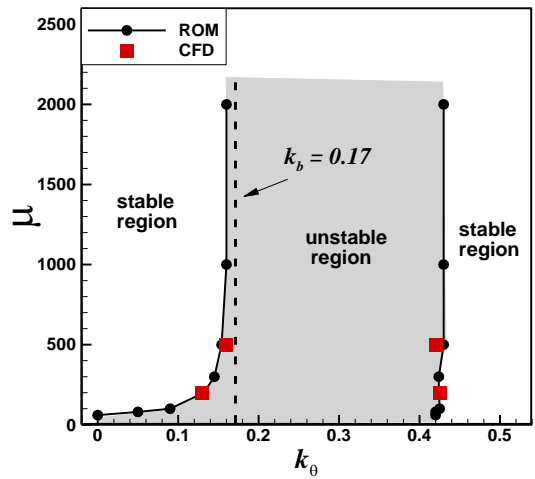

Figure 13. The boot-shaped instability region with different mass ratio at M=0.7 and a=4.5 degrees [11]

Figure 14 shows the Bode diagram as well as its zero and pole of the open-loop system predicted by the ROM in the condition of M=0.7 and $\alpha$=4.5 degrees. It can be seen that the instability region predicted by the phase-frequency curve of the Bode diagram (between 0 degree and 180 degrees) is consistent with those of the ROM-based aeroelastic model shown in Figure 12. The lower boundary (0.17) is determined by the characteristic frequency of the dominant fluid mode. Meanwhile, this boundary corresponds to the maximum value in the amplitude-frequency curve, which represents the pole point of the system; while the upper boundary corresponds to the minimum value, representing the zero point of the system. The most unstable zero and pole points of the system are shown in Figure 14(b), and the characteristic frequencies are in good agreement with the instability boundaries. The instability region of the transonic SDOF flutter (A/B type buzz), therefore, is governed by the frequencies of the zero and pole points of the open-loop system. From this point, the engineer can directly obtain the flutter boundary of interest by a quick analysis from the open-loop model, rather than the inefficient coupling calculation.

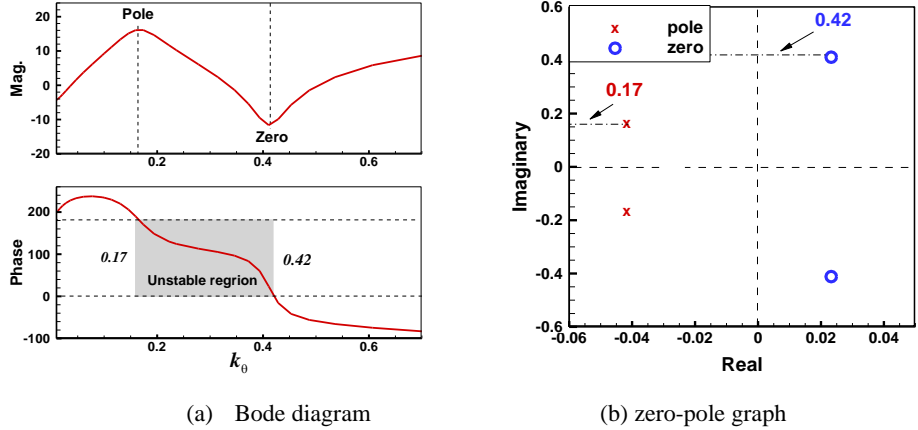

(a) Bode diagram       (b) zero-pole graph

Figure 14. Bode diagram and the zero-pole graph of the open-loop system, M=0.7, a=4.5 degrees

The above conclusion is also applicable to the other states shown in Table 3. Figure 15 shows the root loci of the coupled system at different angles of attack near the buffet onset. It can be found that both the instability region and the strength of the aeroelastic system grow with the increase of the angle of attack (close to the buffet onset angle). This indicates that the occurrence of transonic buzz is closely related to the transonic buffet, or to be more exact, to the stability of the flow. From the root loci, the essence of transonic buzz is the SDOF flutter caused by the coupling between the dominant fluid mode and the structural mode. For this instability to arise, the damping of the flow must be sufficiently low, i.e. the static airfoil is at an angle of attack near the buffet onset. Besides, Figure 16 shows the relationship between the instability boundaries and the frequency corresponding to zero and pole points at different angles of attack. It indicates that the upper and lower instability boundaries are entirely determined by the zero and pole points predicted by the ROM, which proves the physical significance of the frequency boundary to the occurrence of transonic buzz. This conclusion is of significance to understand the dynamics of transonic buzz and to guide the design of the control surface.

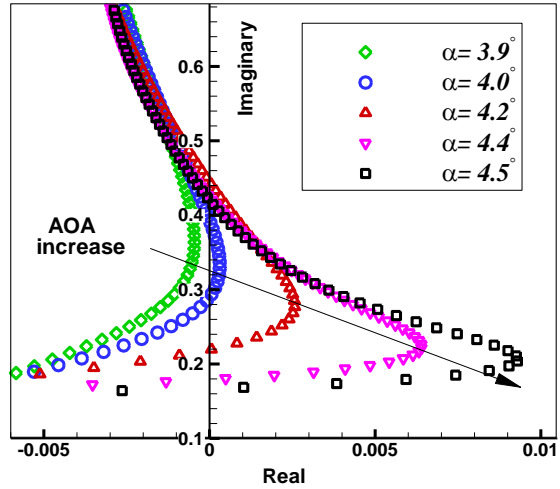

Figure 15 Root loci of the coupled system at different angles of attack [11]

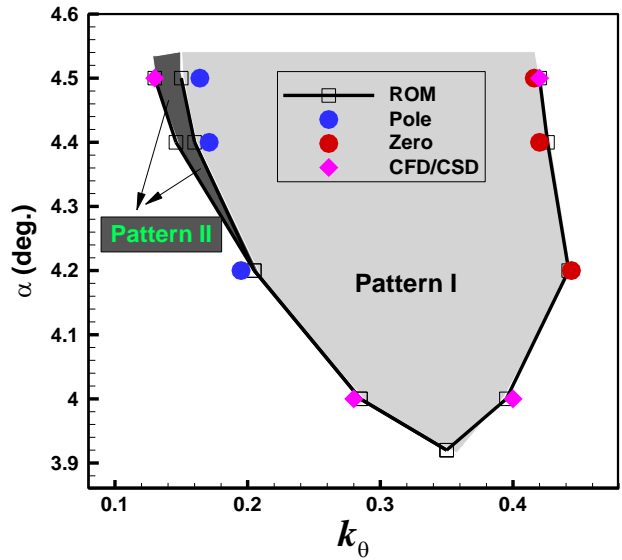

Figure 16. Comparison of the frequency between the instability boundary and the zero-pole point

With the same pitching model and investigated method, the mechanism of C-type buzz is also investigated. The results indicate that the Mach number region for the potential occurrence of C-type buzz is about 0.93 to 1.6, which is almost consistent with the region predicted in the literature [9]. Figure 17 presents the damping (the real part of the eigenvalue) of the dominant fluid mode as a function of the Mach number. Within the potential range, the mode damping is around -0.2. As the Mach number further increases (larger than 1.6), however, the damping of the dominant fluid mode decreases rapidly and transonic buzz will not occur at these Mach numbers. Therefore, the occurrence of C-type buzz is related to the low stability margin of the fluid mode in the low supersonic region. It should also be noticed that this stability margin is obviously higher than that in the transonic region, the potential region for types A and B.

Figure 18 shows the Bode diagram and the zero-pole distribution of the open-loop system in the case of M=1.2. The comparison of instability boundaries between the CFD/CSD simulation and the ROM-based aeroelastic model is shown in Figure 19. It can be seen that the predicted instability boundary of the Bode diagram is consistent with the CFD/CSD simulation. The frequency of the upper boundary is very close to the frequency of the zero point, and the lower boundary is the same as the frequency of the pole point. Similar to A/B type buzz, the instability boundary of C type buzz is also determined by the zero-pole frequency.

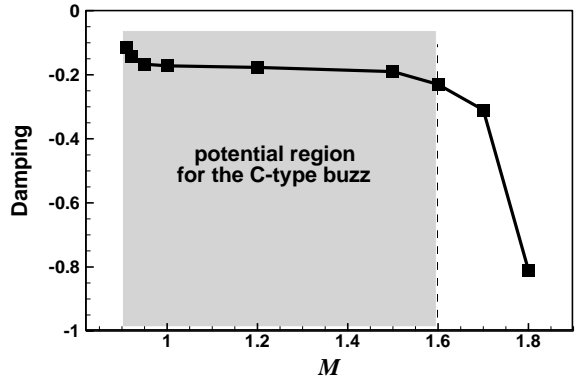

Figure 17 Damping of the dominant fluid mode with different Mach number

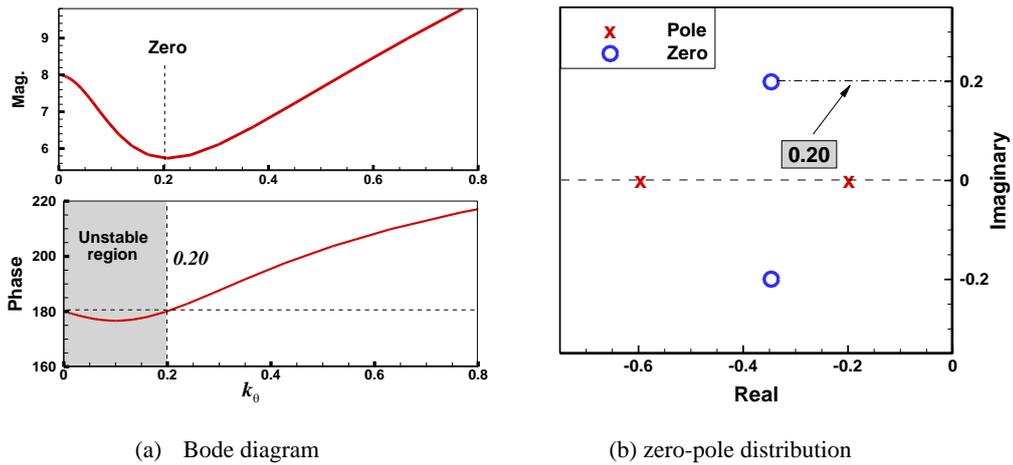

(a) Bode diagram          (b) zero-pole distribution

Figure 18 Bode diagram and the zero-pole distribution of the open-loop system at M = 1.2

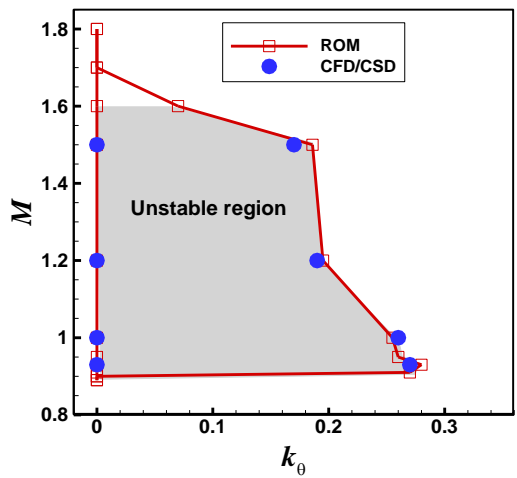

Figure 19 the comparison of the instability boundaries with different method

The results predicted by the ROM-based aeroelastic model are verified by the CFD/CSD simulation. First, transonic buzz is a kind of SDOF flutter. From the above analysis, the present SDOF flutter and the classic bending-torsion flutter can be explained by a unified theory—the mode-coupling theory. But different from the classic flutter which is caused by the coupling between two or more structural modes at high dynamic pressures, the transonic SDOF flutter is the result of the coupling between one structural mode and one dominant fluid mode. Furthermore, two requisite criteria must be fulfilled for the instability to arise. Firstly, the fluid must have a sufficiently low damping, that is, the pre-buffet conditions are close to the onset. Secondly, the structural frequency must be within a certain range, namely, the region between the frequencies of zero and pole of the open-loop system.

From another viewpoint, the SDOF flutter is not a special case that only exists in the transonic flow. As long as the requisite criteria are satisfied, it would also occur in other flow conditions, such as the wing at a high angle of attack [98-100] and the elastically supported blunt body at a low Reynolds number [101-102]. This section provides a new perspective to understand the mechanism of the SDOF flutter when the fluid mode is considered.

## 4.4 Type II: Instability on the fluid mode in the pre-buffet flow (reduction on the transonic buffet onset)

The essence of transonic buzz (the SDOF flutter) discussed in the above section is the instability of the structural mode due to the coupling between the fluid mode and the structural mode when the elastic degree of freedom is activated. This study inspired us to further explore the following questions.

1) Whether the instability in the fluid mode will be provoked in the pre-buffet condition of a wing with activated elasticity? That is, will the buffeting phenomenon occur at a lower angle of attack for an elastic wing?

2) What are the dynamical differences between the flutter-pattern instability and the buffeting-pattern instability?

Answers to these questions are crucial to understand the buffet onset and buffeting loads for an actual elastic wing. Quite recently, Gao et al. [103] investigated the effect of the elastic characteristics on the buffet onset. In their study, a NACA0012 airfoil with activated elasticity in the pitching degree of freedom is adopted. It should be emphasized that the effect of the deformation caused by the static aeroelasticity has been eliminated in advance in the research, which indicates that any changes observed on the stability are not caused by the static aeroelasticity.

For a stationary NACA0012 airfoil, the transonic buffet onset angle is about 4.80 degrees at the Mach number of 0.7. When the freestream angle of attack is lower than the onset angle, the flow is absolutely stable and the time history response is steady. If the pitching degree of freedom is released, however, the system will become unstable in certain combinations of structural parameters due to the FSI effect. Figure 20 shows the real and imaginary parts of the eigenvalue loci of the coupled system as a function of the structural frequency in a typical pre-buffet case of $M = 0.7$, $\alpha = 4.5$ degrees. It can be seen that the eigenvalue loci of interest display two branches, Branch 1 and Branch 2. These branches exchange roles as the structural frequency increases, for example, Branch 1 switches from the fluid mode to the structural mode at $k_\theta = 0.18$. That is, the coupled system exhibits two distinct instability patterns, namely, Pattern I and Pattern II as shown in Figure 20. Pattern I, instability in the structural mode, is in essence the SDOF flutter, the mechanism of which has been discussed in section 4.3.

Different from Pattern I, the dynamics of Pattern II is caused by the instability in the fluid mode (Figure 20a), in which the coupling frequency of the system follows the dominant frequency of the uncoupled fluid mode. It is the transonic buffeting phenomenon under the unstable buffet loads. That is, transonic buffet will be induced at a low angle of attack due to the influence of the FSI when the pitching degree of freedom is activated.

Figure 21 shows the time history responses and the power spectrum density (PSD) results of two typical cases, $k_\theta = 0.13$ in Pattern II and $k_\theta = 0.17$ in Pattern I. At $k_\theta = 0.13$, the response amplitude is far smaller than that of $k_\theta = 0.17$, and the response frequency is 0.182, following the buffet frequency of the rigid airfoil. For comparison, Figure 21 also shows the time history responses and the PSD analysis at the buffet onset angle $\alpha = 4.8$ degrees with the same structural frequency $k_\theta = 0.13$. It is interesting to notice that response amplitudes as well as coupling frequencies in the pre-buffet condition are almost identical with those in the buffet condition. Furthermore, Figure 22 presents the lift coefficient variation with the instantaneous pitching angle. The Lissajou plots in cases of $\alpha = 4.5$ degrees, $k_\theta = 0.13$ and $\alpha = 4.8$ degrees, $k_\theta = 0.13$ are very similar with regard to their shapes and areas. However, they are totally different from those of $\alpha = 4.5$ degrees, $k_\theta = 0.17$. These facts further prove that the dynamics and instability mechanism of Pattern II are significantly different from Pattern I, but they are corresponding to the forced vibration under buffet loads.

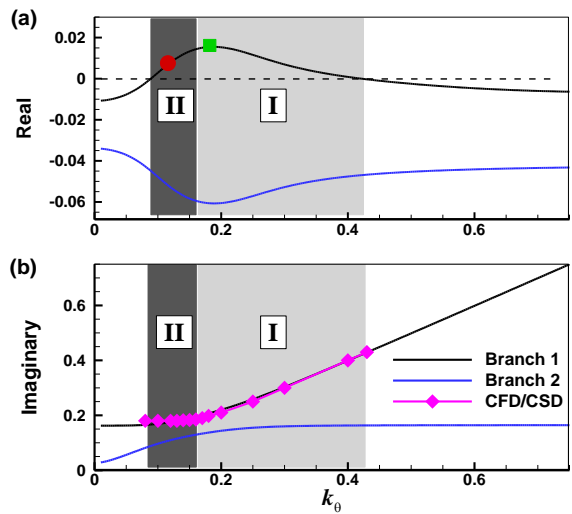

Figure 20 Real and imaginary parts of the eigenvalue loci as a function of the structural frequency [103]

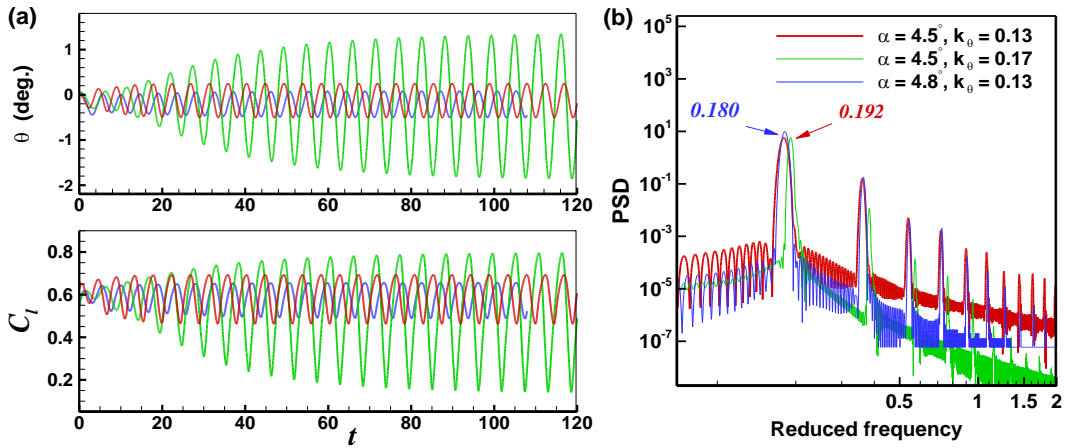

Figure 21 Time history responses and the PSD result of typical cases [103]

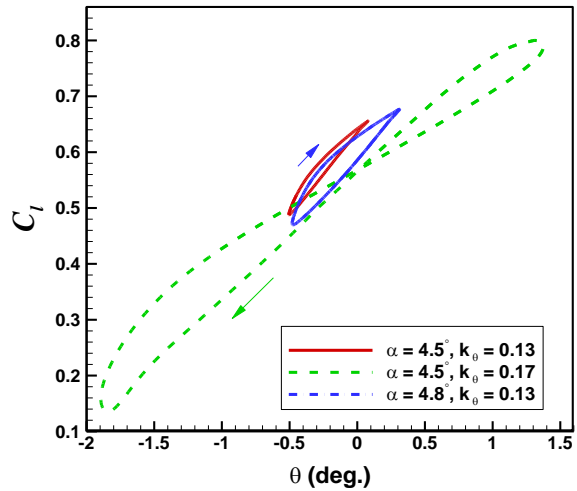

Figure 22 Lift coefficient vs instantaneous pitching angle in typical cases. [103]

The influences of the mass ratio and the structural frequency are further investigated. Figure 23 shows the instability boot with the contour of the frequency ratio (the coupling frequency to the flow characteristic frequency). The right part (the yellow region) indicates the flutter-pattern instability discussed in Section 4.3; while the dynamics in the boot front (the blue region) is governed by the buffeting-pattern instability, in which the frequency ratio is close to 1. In this region, the fluid mode will lose its stability at a lower angle of attack. Figure 24 shows the regions of buffeting-pattern instability and flutter-pattern instability at different pre-buffet angles of attack. At Mach number of 0.7, the lowest angle of attack to induce the buffeting-pattern instability is 4.1 degrees, which is decreased by 0.7 degrees compared with the rigid model. Therefore, elastic characteristics can significantly reduce the onset angle of the transonic buffet, which should be a crucial factor for researchers to investigate transonic buffet for a real elastic wing.

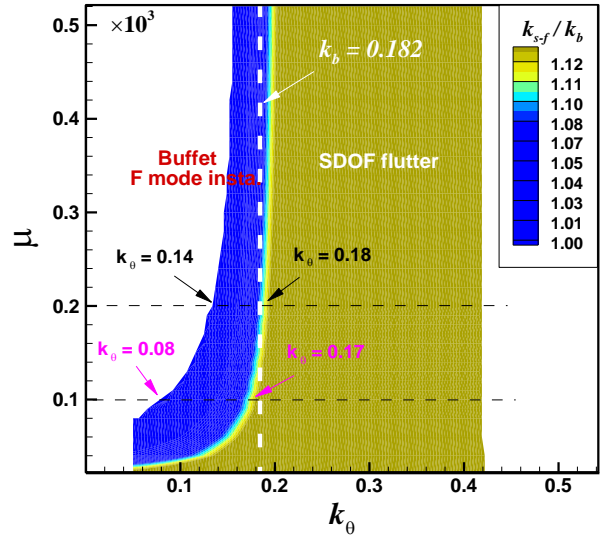

Figure 23 The instability boot as a function of the mass ratio and the reduced structural frequency [103]

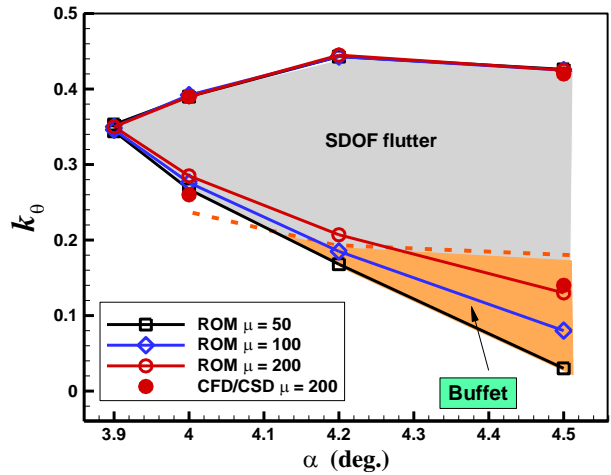

Figure 24 Regions of buffeting-pattern instability and flutter-pattern instability in different pre-buffet conditions

## 4.5 Type III: Forced vibration in unstable flows (buffeting response)

Transonic buffeting refers to the forced vibration of aircraft structures due to the excitation of unstable separated flows. The vibration will not only affect the fatigue life of an aircraft structure, but also the operation of the instrumentation equipment and the use of the flight control system.

The research process of the transonic buffeting phenomenon in industry has been divided into two uncoupled steps: predicting the buffet loads based on the rigid wing first and then calculating the response of the real elastic wing under the given loads. This research route ignores the interaction between the fluid and the structure. The cost and difficulty of the coupling research are much higher than those of the uncoupled method in both experiment and numerical computation. In the uncoupled framework, the objective is to predict the buffet load. Since the buffet flow itself is independent from the motion of the wing, it is often regarded as a pure fluid mechanics problem. In this field, transonic buffet is a phenomenon of flow global instability, characterized by the periodic low-frequency and large-amplitude shock oscillations, at a certain combination of Mach number and mean angle of attack.

Following this viewpoint, most studies on transonic buffet were conducted using the rigid and stationary wing/airfoil. These studies focused on the buffet mechanism, buffet onset and loads prediction, as well as buffet flow control. The research on the buffet mechanism aims to reveal the root cause of the large-amplitude shock oscillation by wind tunnel test and numerical simulation. The self-excited feedback model proposed by Lee [104-105], and the global instability model proposed by Crouch [106-107] are the most successful explanations. Buffet onset prediction is to find the combination of Mach number and angle of attack [28,108]. For the civil aircraft, a certain margin must be reserved between the cruise state and the buffet onset boundary. The accurate prediction of buffet onset, therefore, is a key in the design process. Buffet load prediction is another important task for the aircraft engineers [31,109]. As mentioned before, most of the current numerical studies focus on how to achieve the high-precision buffeting response by investigating the sensitivity of simulations to turbulence modelling, spatial and temporal discretization and numerical schemes. The purpose of buffet control is to reduce or even eliminate the unsteadiness of buffet flow as much as possible through appropriate control strategies, thereby reducing the vibration level of the wing [110-114].

A more comprehensive introduction to transonic buffeting can be found in the relevant literature review. Bendiksen [2] offered an overview of the development and application of the unsteady transonic flow theory, including a brief review of transonic buffeting. Lee [105] summarized the research results of transonic airfoil buffeting before 2000, focusing on the feedback model of shock-boundary layer interference. Diannelis [115] provided a comprehensive overview of the current research progress of transonic buffeting, especially the research achievements in the past ten years, including the new understanding of buffeting mechanism, the new progress in numerical simulation, the new discovery of dynamic response in buffeting flow and the research and understanding of three-dimensional buffeting flow.

## 4.6 Type IV: Instability on the structural mode in the unstable flow (lock-in in transonic buffet flow)

In classical aeroelasticity, the wing would exhibit a forced vibration due to the unstable buffet flow. In this framework, the wing vibrating frequency should depend on the buffet frequency. However, it is found that it no longer follows the buffet frequency but instead, locks onto the natural frequency of the wing when the natural frequency approaches the buffet frequency (Figure 25). Simultaneously, a large oscillating amplitude of the airfoil is observed within the lock-in region, which is dangerous to the aircraft. This abnormal phenomenon is referred to as "frequency lock-in" in transonic buffet flow. This phenomenon, for a long time, has been believed to be the result of the nonlinear aerodynamic resonance, for instance, the research by Raveh [17-18] and Hartmann [19]. However, this interpretation has its limitations [116]. On one hand, the lock-in region did not display a symmetrical distribution against the buffet frequency as expected by the resonance theory. As shown in Figure 25, the lock-in phenomenon still existed when the frequency ratio was enlarged to $k_\theta / k_b \sim 2.4$. On the other hand, the maximum vibration amplitude of the structure was obtained at the frequency ratio $k_\theta / k_b \sim 1.5$ rather than at the synchronized point $k_\theta / k_b \sim 1.0$. These anomalies are difficult to explain by the perspective of resonance. Then, what is the real physical mechanism of the frequency lock-in phenomenon? And what is the consequence of the oscillation with a large amplitude in transonic buffet flow?

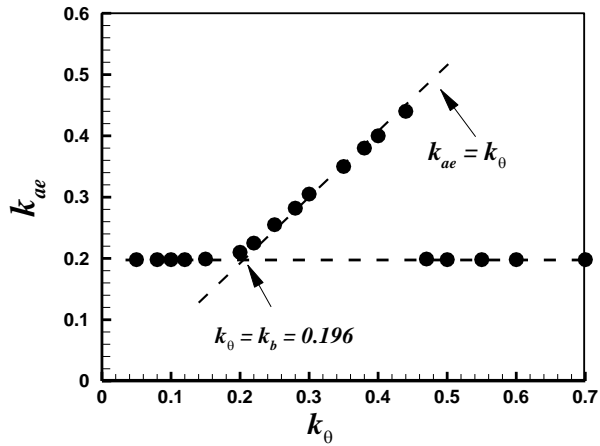

Figure 25 Coupling frequency of the aeroelastic system as a function of the natural frequency of the airfoil [20]

Very recently, Gao et al. [20] revealed the mechanism of frequency lock-in in transonic buffet flow through the ROM-based aeroelastic model and CFD/CSD simulation. In their study, the investigated model is also the spring suspended NACA0012 airfoil. The flow condition is fixed at M=0.7, a=5.5 degrees and Re=3×10⁶, a post buffet condition with the strongest buffet load. And in this case, the buffet frequency $k_b$ is 0.196. Figure 26 shows the real part and imaginary part of the eigenvalue loci, as well as the coupling frequency and the amplitude of the structural response obtained by the CFD/CSD simulation as a function of the structural frequency. In subregion 3, the damping (the real part) of structural mode ($S$ mode) is positive (Figure 26b), which indicates the instability in this mode, namely, flutter in aeroelasticity. The instability range coincides with the lock-in region obtained from the coupled CFD/CSD simulation (Figure 26c). It reveals that the frequency lock-in is caused by the linear coupled-mode flutter. But different from the SDOF flutter discussed in Section 4.3, the present flutter case is induced in the unstable buffet flow. The unstable flow condition may mislead the researchers to understand this problem as a forced vibration. However, Gao's [20] research reveals that flutter is the primary cause of the frequency lock-in phenomenon in transonic buffet flow. And based on this mechanism, the limitations in the resonance interpretation can be well explained.

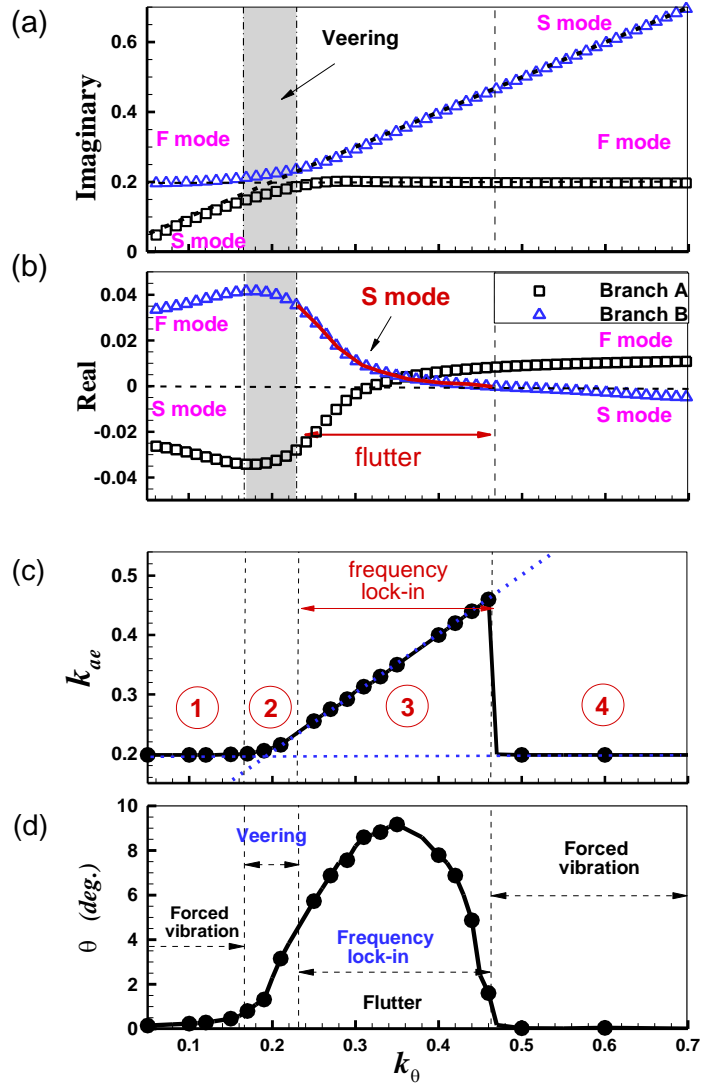

Figure 26 (a) the imaginary part of the eigenvalue loci, (b) the real part of the eigenvalue loci from the ROM-based aeroelastic model; and (c) coupling frequency of the system, (d) oscillating amplitude of the pitching airfoil from the coupled CFD/CSD simulation varying with the natural frequency of the elastic airfoil [20]

The dynamic characteristics of the spring suspended airfoil in transonic buffet flow can be divided into four patterns, as shown in Figure 26c. In the first pattern (with a small structural frequency, $k_\theta < 0.168$) and the fourth pattern (with a high structural frequency, $k_\theta > 0.460$), the coupling frequencies follow the buffet flow frequency, namely, $k_{ae} = k_b$. In fact, they are both forced vibration caused by the unsteady buffeting loads, in which the vibration amplitudes of the airfoil are comparatively small. In the third pattern (with a medium structural frequency, $0.232 < k_\theta < 0.460$), the coupling frequency synchronizes with the natural frequency ($k_{ae} = k_\theta$) — the lock-in pattern caused by flutter. The peak oscillation amplitude is achieved at a distinctly high frequency ratio of $k_\theta / k_b = 1.73$. The second is a transitional pattern ($0.168 < k_\theta < 0.232$), namely, the mode veering region. It is hard to identify which mode, $S$ mode or $F$ mode, dominates the system instability. This pattern will not exist when the mass ratio is relatively high, i.e. 1000 in Gao's study [20].

In the lock-in pattern, there is a competition between two unstable modes, the induced unstable structural mode and the original unstable fluid mode. Therefore, the response of the system undergoes a **conversion** from the forced vibration to the self-sustained oscillation (flutter). Figure 27 shows the conversion process obtained by the CFD/CSD simulation at $\mu = 200$ and $k_\theta = 0.3$. At the beginning, $t < 200$, the coupling frequency is followed by the buffet frequency ($k_b$=0.196) [Figure 27c]. The system is dominated by the unstable fluid mode (**F** mode), displaying the forced vibration under the unstable buffet loads. When $200 < t < 440$, there are two peak frequency components, the buffet frequency with less power, and the structural frequency component with more power [figure 27d]. It means that, as the response time increases, the buffet frequency becomes weak, while the structural frequency becomes strong. That is, the unstable **S** mode gradually dominates the dynamic characteristics of the system. Finally, when $t > 440$, the responses diverge, and the oscillating frequency absolutely locks-onto the structural frequency [Figure 27e]. The frequency lock-in phenomenon occurs. The unstable **S** mode completely dominates the characteristics of the system.

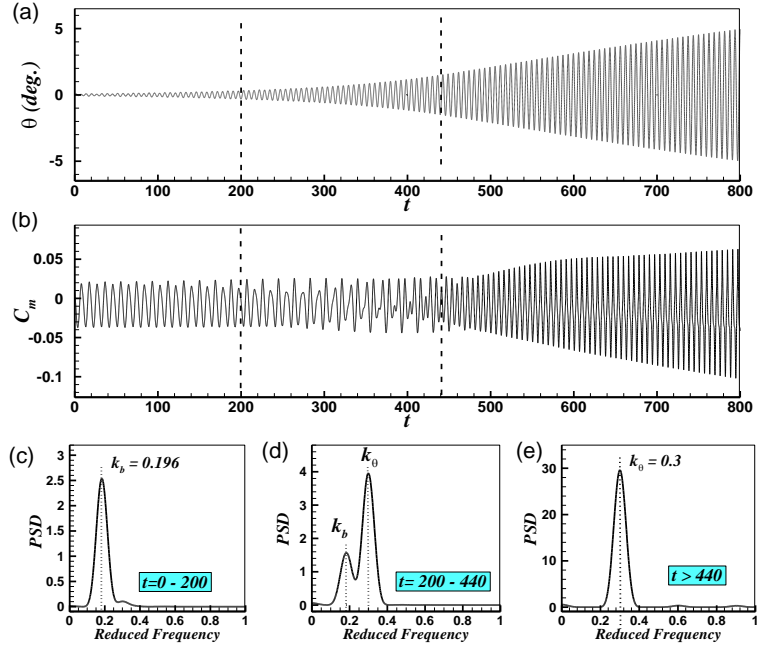

Figure 27 The conversion process of the system in the lock-in pattern [20]

The large-amplitude structural vibration in the frequency lock-in region may cause flight accidents. However, there is a significant difference in the predicted amplitude between coupled and uncoupled methods. Figure 28 shows the response amplitudes as a function of the structural frequency by both coupled method and uncoupled method — the two-step approach discussed in Section 4.5. In the uncoupled method, fairly low structural damping ratios, 0.5% and 1%, are considered. As can be seen from Figure 28, the maximum amplitude is obtained at the resonance point by the uncoupled method, but it is still one order of magnitude smaller than that of the coupled method.

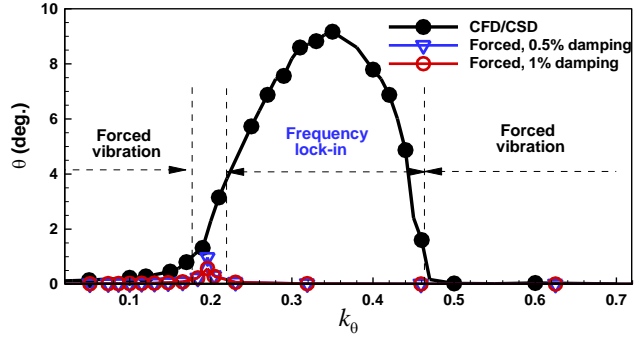

Figure 28 Comparison of vibration amplitude obtained by the CFD/CSD simulation and the uncoupled method

The above research not only provides a reasonable explanation of the frequency lock-in phenomenon, but also overturns the traditional uncoupled approach for the transonic buffet problem in aircraft design. The large-amplitude vibration is essentially the result of the coupling effect between the unstable flow and the structural motion. On the contrary, the results obtained by uncoupled method failed to predict the maximum amplitude and the corresponding structural frequency. This research is valuable to the structural design and the vibration control in the field of aeronautical engineering.

# 5. Concluding remarks

From the perspective of flow stability, this paper discusses the coupled patterns between the structural mode and the fluid mode by a low-order aeroelastic model, and provides a systematic review on the complexity and mechanism of aeroelastic problems associated with transonic buffet. The selected cases include the transonic single-degree-of-freedom flutter (or referred as transonic buzz), the reduction of transonic buffet onset, the classical transonic buffeting response problem and the frequency lock-in phenomenon in the transonic buffet flow. Main conclusions can be summarized as follows:

1)  The introduction of the fluid mode provides a new perspective to understand transonic aeroelastic phenomena. From this perspective, the physical mechanisms of different aeroelastic phenomena are clearly revealed in a unified framework. Due to the decrease of flow stability, the pole representing the fluid mode is added to the fluid-structure interaction equation, and the fluid mode assumes a main role in the coupling process, resulting in different instability patterns as well as different aeroelastic phenomena. The dimensionless frequency ratio between the structural mode and the fluid mode is a key parameter to affect the characteristics of the coupled system.

2)  It clarifies the difference and relationship among these transonic aeroelastic phenomena from the instability patterns. The reduction of transonic buffet onset (Type II) and the classical transonic buffeting response (Type III) are both caused by the instability in the fluid mode. The response amplitude of the structure is small, and the response frequency is consistent with the unstable flow frequency. However, the nature of transonic buzz (Type I) and the frequency lock-in phenomenon (Type IV) are the instability in the structural mode. In these flutter cases, the response amplitudes are relatively higher than those in the forced vibration case, and the response frequency locks onto the natural frequency of the structure. Due to the nonlinearity of transonic flow, these flutter cases often reveal themselves in the form of limit cycle oscillations.

3) The complexity of transonic aeroelastic problems lies in the decrease and even instability of the flow. The unstable freestream not only complicates the coupling process, but also causes a misleading of contributing these aeroelastic phenomena to the forced vibration. However, in fact, most large-amplitude vibration problems in separated flows are still dominated by the self-excited flutter, especially for the frequency lock-in phenomenon. For the case with a strong fluid-structure feedback effect, the dynamics is mainly caused by the instability in the structural mode, which is essentially a flutter problem. The damage caused by this kind of problem is often more serious than that caused by the resonance problem.

# 6. Acknowledgements

This work was supported by the National Natural Science Foundation of China (Grant No. 11622220 and 11902269), the 111 Project of China (B17037), and the Innovation Foundation for the Postdoctoral Talents (Grant No. BX20180258), and China Postdoctoral Science Foundation funded project (No 2019M653725). This work was also supported by Yong Talent fund of University Association for Science and Technology in Shaanxi, China.

**Reference**

[1] Badcock K J, Timme S, Marques S, et al. Transonic aeroelastic simulation for instability searches and uncertainty analysis. Progress in Aerospace Sciences, 2011, 47(5):392-423.

[2] Bendiksen O O. Review of unsteady transonic aerodynamics: theory and applications. Progress in Aerospace Sciences, 2011, 47(2): 135-167.

[3] Dowell E H. Some recent advances in nonlinear aeroelasticity: fluid-structure interaction in 21st Century. In: 51st AIAA/ASME/ASCE/AHS/ASC Structures, Structural Dynamics, and Materials Conference, 2010.

[4] Yates E C, AGARD standard aeroelastic configurations for dynamic response I-wing 445.6. AGARD Report No. 765, 1988.

[5] Isogai K. Transonic dip mechanism of flutter of a sweptback wing: Part II. AIAA Journal, 1981, 19(9): 1240-1242.

[6] Mallik W ,Schetz J A , Kapania R K . Rapid Transonic Flutter Analysis for Aircraft Conceptual Design Applications. AIAA Journal, 2018, 56(6): 2389-2402.

[7] Opgenoord M MJ ,Drela M , Willcox K E . Physics-based low-order model for transonic flutter prediction. AIAA Journal, 2018,56(4):1519-1531.

[8] Silva W A,Lii B P, ChwalowskiP. Evaluation of linear, inviscid, viscous, and reduced-order modelling aeroelastic solutions of the AGARD 445.6 wing using root locus analysis. International Journal of Computational Fluid Dynamics, 2014, 28(3): 122-139.

[9] Lambourne N C. Control-surface buzz. In: HM Stationery Office, Reports and Memoranda No. 3364, 1964.

[10] Bendiksen O O. Nonclassical aileron buzz in transonic flow. In 34[th] AIAA/ASME/ASCE/AHS/ASC Structures, Structural Dynamics and Meterials Conference. La Jolla, CA: American Institute of Aeronautics and Astronautics, AIAA 93-1479, 1993.

[11] Gao C Q, ZhangW W, Ye Z Y. A new viewpoint on the mechanism of transonic single-degree-of-freedom flutter. Aerospace Science and Technology, 2016, 52, 144-156.

[12] Gao C Q., Zhang W W, Liu Y L, et al.Numerical study on the correlation of transonicsingle-degree-of-freedom flutter and buffet.Science China Physics, Mechanics & Astronomy, 2015, 58(8), 084701.

[13] Steimle C, Karhoff D C, Schröder W. Unsteady transonic flow over a transport-type swept wing. AIAA Journal, 2012, 50(2): 399-415.

[14] Zhang W W, Gao C Q, Liu Y L, et. al. The interaction between flutter and buffet in transonic flow. Nonlinear Dynamics,2015, 82(12): 1851-1865.

[15] Chiarelli M R, Bonomo S. Numerical investigation into flutter and flutter-buffet phenomena for a swept wing and a curved planform wing. International Journal of Aerospace Engineering, 2019, 8210235.

[16] Kou J Q, Zhang W W. A hybrid reduced-order framework for complex aeroelastic simulations. Aerospace Science and Technology, 2019, 84: 880-894.

[17] Raveh, D. E. & Dowell, E. H. Frequency lock-in phenomenon for oscillating airfoils in buffeting flows. Journal of Fluids and Struct. 2011, 27 (1), 89–104.

[18] Raveh DE, Dowell E H. Aeroelastic responses of elastically suspended airfoil systems in transonic buffeting flows . AIAA Journal, 2014, 52 (5): 926–934.

[19] Hartmann A, Klaas M, Schröder W. Coupled airfoil heave/pitch oscillations at buffet flow . AIAA Journal, 2013, 51 (7): 1542–1552.

[20] Gao C Q, Zhang W W, Li X T, et al. Mechanism of frequency lock-in in transonic buffeting flow . Journal of Fluid Mechanics, 2017, 818, 528-561.

[21] Parker E C, Spain C V, Soistmann D L. Aileron buzz investigated on several generic NASP wing configurations. In 32[nd] Structures, Structural Dynamics, and Materials Conference. Baltimore MD U.S.A.: AIAA, AIAA-91-0936-CP, 1991.

[22] Acar P , Nikbay M . Steady and Unsteady Aeroelastic Computations of HIRENASD Wing for Low and High Reynolds Numbers. 54th AIAA/ASME/ASCE/AHS/ASC Structures, Structural Dynamics, and Materials Conference, April 8-11, 2013, Boston, Massachusetts, USA. 2013.

[23] Rivera A J, Dansberry B E, Bennett R M, NACA0012 benchmark model experimental flutter results with unsteady pressure distributions, NASATM 107581, 1992.

[24] Steger J L, Bailey H E. Calculation of transonic aileron buzz. AIAA Journal, 1980, 18(3): 249-255.

[25] Hall K C, Thomas J P, Dowell E H. Proper orthogonal decomposition technique for transonic unsteady aerodynamic flows. AIAA Journal, 2000, 38(10): 1853-1862

[26] Lucia D J, Beran P S, Silva W A. Reduced-order modeling: new approaches for computational physics. Progress in Aerospace Sciences, 2004, 40(1): 51-117.

[27] Pearcey H H. A method for the prediction of the onset of buffeting and other separation effects from wind tunnel

tests on rigid models. Reston: AGARD-Report-223, 1958.

[28]McDevitt JB, Okuno AF. Static and dynamic pressure measurements on a NACA 0012 airfoil in the Ames HighReynolds Number Facility. NASA TP-2485. Reston: NationalAeronautics and Space Administration, 1985.

[29]Lee B H K. Oscillatory shock motion caused by transonic shock boundary-layer interaction. AIAA Journal, 1990, 28(5): 942-944.

[30]Doerffer P, Hirsch C, Dussauge J P, et al. NACA0012 with aileron. Unsteady Effects of Shock Wave Induced Separation. Springer Berlin Heidelberg, 2011: 101-131.

[31]Jacquin L, Molton P, Deck S, et al. Experimental study of shock oscillation over a transonic supercritical profile. AIAA Journal, 2009, 47(9): 1985-1994.

[32]Hartmann A, Feldhusen A, Schröder W. On the interaction of shock waves and sound waves in transonic buffet flow. Physics of Fluids, 2013, 25(2): 89-101.

[33]Iovnovich M, Raveh D E. Numerical study of shock buffet on three-dimensional wings. AIAA Journal, 2014, 53(2): 449-463.

[34]Dandois J. Experimental study of transonic buffet phenomenon on a 3D swept wing. Physics of Fluids, 2016, 28(1): 016101.

[35]Nixon D. An analytic model for control surface buzz. AIAA 98-0417, 1998.

[36]Govardhan R N, Williamson C H K. Resonance forever: existence of a critical mass and an infinite regime of resonance in vortex-induced vibration. Journal of Fluid Mechanics, 2002, 473, 147–166.

[37]Dowell E H, Tang D. Nonlinear aeroelasticity and unsteady aerodynamics. AIAA Journal, 2002, 40(9): 1697-1707.

[38]Dowell E, Edwards J, Strganac T. Nonlinear aeroelasticity. Journal of Aircraft, 2003, 40(5): 857-874.

[39]Afonso F, Vale J, Oliveira E. A review on non-linear aeroelasticity of high aspect-ratio wings. Progress in Aerospace Science, 2017, 89: 40-57.

[40]Schewe G, Mai H, Dietz G. Nonlinear effects in transonic flutter with emphasis on manifestations of limit cycle oscillation . Journal of Fluids and Structures, 2003, 18(1): 3-22.

[41]Jack R E. Numerical simulations of shock/boundary layer interactions using time dependent modeling techniques: a survey of recent results . Progress in Aerospace Sciences, 2008, 44(6): 447-465.

[42]Deck S. Numerical simulation of transonic buffet over a supercritical airfoil. AIAA Journal, 2005, 43(7): 1556-1566.

[43]Chen L W, Xu C Y, Lu X Y. Numerical investigation of the compressible flow past an aerofoil . Journal of Fluid Mechanics, 2010, 643: 97-126.

[44]Huang J B, Xiao Z X, Liu J, et al. Simulation of shock wave buffet and its suppression on an OAT15A supercritical airfoil by IDDES. Science China Physics, Mechanics and Astronomy, 2012, 55(2): 260-271.

[45]Grossi F, Braza M, Hoarau Y. Delayed Detached-Eddy Simulation of the transonic flow around a supercritical airfoil in the buffet regime. Progress in Hybrid RANS-LES Modelling, 2012: 369-378.

[46]Grossi F. Braza M, Hoarau Y. Prediction of transonic buffet by delayed detached-eddy simulation. AIAA Journal, 2014, 52 (10): 2300–2312.

[47]Sartor F, Timme S. Delayed detached-Eddy simulation of shock buffet on half wing-body configuration. AIAA

Journal, 2017, 55(4): 1230-1240.

[48] Barakos G, Drikakis D. Numerical simulation of transonic buffet flows using various turbulence closures. International Journal of Heat and Fluid Flow, 2000, 21(5): 620-626.

[49] Goncalves E, Houdeville R. Turbulence model and numerical scheme assessment for buffet computations. International Journal for Numerical Methods in Fluids, 2004, 46(11): 1127-1152.

[50] Thiery M, Coustols E. Numerical prediction of shock induced oscillations over a 2D airfoil: Influence of turbulence modelling and test section walls. International Journal of Heat and Fluid Flow, 2006, 27(4): 661-670.

[51] Kourta A, Petit G, Courty J C, et al. Buffeting in transonic flow prediction using time-dependent turbulence model. International Journal for Numerical Methods in Fluids, 2005, 49(2): 171-182.

[52] Xiao Q, Tsai H M, Liu F. Numerical study of transonic buffet on a supercritical airfoil. AIAA Journal, 2006, 44(3): 620-628.

[53] Plante F, Laurendeau E. Simulation of transonic buffet using a time-spectral method. AIAA Journal, 2019, 57(3): 1275-1287.

[54] Drofelnik J, Da Ronch A, Franciolini M, Crivellini A. Fast identification of transonic buffet envelope using computational fluid dynamics. Aircraft Engineering and Aerospace Technology, 2019, 91(2): 309-316.

[55] Sartor F, Mettot C, Sipp D. Stability, receptivity, and sensitivity analyses of buffeting transonic flow over a profile. AIAA Journal, 2015, 53(7): 1980–1993.

[56] Sartor F, Mettot C, Bur R, et al. Unsteadiness in transonic shock-wave/boundary-layer interactions: experimental investigation and global stability analysis. Journal of Fluid Mechanics, 2015, 781(1):550-577.

[57] Gao C Q, Zhang W W, Ye Z Y. Numerical study on closed-loop control of transonic buffet suppression by trailing edge flap. Computers and Fluids, 2016, 132, 32-45.

[58] Soda A, Ralph V. Analysis of transonic aerodynamic interference in the wing-nacelle region for a generic transport aircraft. In: Proceedingsof IFSAD 2005 - International Forum on Aeroelasticity andStructural Dynamics. Munich: DLR, 2005.

[59] Dowell E H, Hall K C. Modeling of fluid-structure interaction. Annual Review of Fluid Mechanics, 2001, 33(1): 445-490.

[60] Hall K C, Thomas J P, Dowell E H. Proper orthogonal decomposition technique for transonic unsteady aerodynamic flows. AIAA Journal, 2000, 38(10): 1853-1862

[61] Amsallem, D., Farhat, C., 2008. Interpolation method for adapting reduced-order models and application to aeroelasticity. AIAA Journal. 46 (7), 1803-1813.

[62] Couplet M, Basdevant C, Sagaut P. Calibrated reduced-order POD-Galerkin system for fluid flow modelling. Journal of Computational Physics, 2005, 207(1): 192-220.

[63] Chen G, Sun J, Li Y. Active flutter suppression control law design method based on balanced proper orthogonal decomposition reduced order model. Nonlinear Dynamics, 2012, 70(1): 1-12.

[64] Zhou Q, Chen G, Da Ronch A, Li Y M. Reduced order unsteady aerodynamic model of a rigid aerofoil in gust encounters. Aerospace Science and Technology, 2017, 63: 203-213.

[65] Walton S, Hassan O, Morgan K. Reduced order modelling for unsteady fluid flow using proper orthogonal decomposition and radial basis functions. Appl. Math. Model. 2013, 37 (20-21), 8930-8945.

[66] Schmid P J. Dynamic mode decomposition of numerical and experimental data. Journal of fluid mechanics, 2010, 656: 5-28.

[67] Rowley C W, Mezić I, Bagheri S, et al. Spectral analysis of nonlinear flows. Journal of fluid mechanics, 2009, 641: 115-127.

[68] Sayadi T, Schmid P J, Richecoeur F, et al. Parametrized data-driven decomposition for bifurcation analysis, with application to thermo-acoustically unstable systems. Physics of Fluids, 2015, 27(3):030801.

[69] Kou J, Zhang W. An improved criterion to select dominant modes from dynamic mode decomposition. European Journal of Mechanics - B/Fluids, 2017, 62:109-129.

[70] Thomas J P, Dowell E H, Hall K C. Nonlinear inviscidaerodynamic effects on transonic divergence, flutter and limit cycle oscillations. AIAA Journal, 2002, 40(4): 638-646.

[71] Da Ronch, A., Mccracken, A. J., Badcock, K. J., Widhalm, M., Campobasso, M. S., 2013. Linear Frequency Domain and Harmonic Balance Predictions of Dynamic Derivatives. Journal of Aircraft, 50 (3): 694-707.

[72] Da Ronch A, Badcock K J, Wang Y, Wynn A, Palacios R. Nonlinear model reduction for flexible aircraft control design. In: AIAA Atmospheric Flight Mechanics Conference. 2012, AIAA 2012-4404.

[73] Silva W. Identification of Nonlinear Aeroelastic Systems Based on the Volterra Theory: Progress and Opportunities. Nonlinear Dynamics, 2005, 39(1-2):25-62.

[74] Yao W, Jaiman R K. Model reduction and mechanism for the vortex-induced vibrations of bluff bodies. Journal of Fluid Mechanics, 2017, 827: 357-393.

[75] Zhang W W, Ye Z Y. Reduced-order-model-based flutter analysis at high angle of attack. Journal of Aircraft, 2007, 44(6):2086-2089.

[76] Zhang W, Ye Z, Zhang C. Aeroservoelastic Analysis for Transonic Missile Based on Computational Fluid Dynamics. Journal of Aircraft, 2009, 46(6):2178-2183.

[77] Zhang W W, Li X T, Ye Z Y, Jiang Y W. Mechanism of frequency lock-in in vortex-induced vibrations at low Reynolds numbers. Journal of Fluid Mechanics, 2015, 783: 72-102.

[78] Zhang W W, Chen K J, Ye Z Y. Unsteady aerodynamic reduced-order modeling of an aeroelastic wing using arbitrary mode shapes. Journal of Fluids and Structures, 2015, 58: 254-270.

[79] Chen G, Li D F, Zhou Q, Da Ronch A, Li Y M. Efficient aeroelastic reduced order model with global structural modifications. Aerospace Science and Technology, 2018, 76: 1-13.

[80] Marzocca P, Librescu L, Silva W. Nonlinear Open/Closed-Loop Aeroelastic Analysis of Airfoils via Volterra Series. AIAA Journal, 2004, 42(4):673-686.

[81] Huang R, Hu H, Zhao Y. Nonlinear reduced-order modeling for multiple-input/multiple-output aerodynamic systems. AIAA Journal, 2014, 52(6): 1219-1231.

[82] Kou J Q, Zhang W W, Yin M L. Novel Wiener models with a time-delayed nonlinear block and theiridentification. Nonlinear Dynamics, 2016, 85(4): 2389-2404.

[83] Kou J Q, Zhang W W. Layered reduced-order models for nonlinear aerodynamics and aeroelasticity. Journal of Fluids and Structures, 2017, 68: 174-193.

[84] Kou J Q, Zhang W W. Multi-kernel neural networks for nonlinear unsteady aerodynamic reduced-order modeling. Aerospace Science and Technology, 2017, 67: 309-326.

[85] Winter M, Breitsamter C. Neurofuzzy-model-based unsteady aerodynamic computations across varying freestream conditions. AIAA Journal, 2016, 54(9): 2705-2720.

[86] Winter M, Breitsamter C. Nonlinear identification via connected neural networks for unsteady aerodynamic analysis. Aerospace Science and Technology, 2018, 77:802-818.

[87] Winter M, Breitsamter C. Coupling of recurrent and static neural network approaches for improved multi-step ahead time series prediction. In Dillmann et al., editor, New Results in Numerical and Experimental Fluid Mechanics XI, Springer-Verlag, 2018, pp 433–442.

[88] Li K, Kou J Q, Zhang W W. Deep neural network for unsteady aerodynamic and aeroelastic modeling across multiple Mach numbers. Nonlinear Dynamics, 2019: 1-21.

[89] Zhang W, Jiang Y, Ye Z. Two Better Loosely Coupled Solution Algorithms of CFD Based Aeroelastic Simulation. Engineering Applications of Computational Fluid Mechanics, 2007, 1(4):253-262.

[90] Dowell E H et al. A modern course in aeroelasticity. Dordrecht: Kluwer Academic Publishers, 4th edition. 2004.

[91] Kou J Q, Zhang W W, Liu Y L, Li X T. The lowest Reynolds number of vortex-induced vibrations. Physics of Fluids, 2017, 29: 041701.

[92] Erickson A L, Stephenson J D. A suggested method of analyzing for transonic flutter of control surfaces based on available experimental evidence. Technical Report Archive & Image Library, NACA RM A7F30, 1947.

[93] W.H. Phillips, J.J. Adams, Low-speed tests of a model simulating the phenomenon of control-surface buzz, NACA RM L50F19, 1950.

[94] J.D. Lang, A model for the dynamics of a separation bubble used to analyze control-surface buzz and dynamic stall, in: AIAA Paper 75-867, 1975.

[95] Bendiksen O O. Nonclassical aileron buzz in transonic flow. In: Proceedings of 34th Structures, Structural Dynamics and Materials Conference. La Jolla: AIAA, 1993

[96] He S, Yang Z, Gu Y. Limit cycle oscillation behavior of transonic control surface buzz considering free-play nonlinearity. Journal of Fluids and Structures, 2016, 61: 431–449.

[97] Gao C Q, Zhang W W. Study on the mechanism and instability parameters of transonic buzz. Acta Aerodynamica Sinica, 2019, 37(1): 99-106. (in Chinese)

[98] Dimitriadis G, Li J. Bifurcation behavior of airfoil undergoing stall flutter oscillations in low-speed wind tunnel. AIAA Journal, 2009, 47(11): 2577-2596.

[99] Bhat S S, Govardhan R N. Stall flutter of NACA 0012 airfoil at low Reynolds Numbers. Journal of Fluids and Structures, 2013, 41: 166-174.

[100] Poirel D, Goyaniuk L, Benaissa A. Frequency lock-in in pitch–heave stall flutter. Journal of Fluids and Structures, 2018, 79: 14-25.

[101] Mittal S, Singh S. Vortex-induced vibrations at subcritical Re. Journal of Fluid Mechanics, 2005, 534:185-194.

[102] de Langre E. Frequency lock-in is caused by coupled-mode flutter. Journal of Fluids and Structures, 2006, 22, 783-791.

[103] Gao C Q, Zhang W W, Ye Z Y. Reduction of transonic buffet onset for a wing with activated elasticity. Aerospace Science and Technology, 2018, 77: 670-676.

[104]    Lee B H K. Oscillatory shock motion caused by transonic shock boundary-layer interaction. AIAA Journal, 1990, 28(5): 942-944.

[105]    Lee B H K. Self-sustained shock oscillations on airfoils at transonic speeds. Progress in Aerospace Sciences, 2001, 37(2): 147-196.

[106]    Crouch J D, Garbaruk A, Magidov D. Predicting the onset of flow unsteadiness based on global instability. Journal of Computational Physics, 2007, 224(2): 924-940.

[107]    Crouch J D, Garbaruk A, Magidov D, et al. Origin of transonic buffet on aerofoils. Journal of Fluid Mechanics, 2009, 628: 357-369.

[108]    Yang Z C, Dang H X. Buffet onset prediction and flow field around a buffeting airfoil at transonic speeds. AIAA-2010-3051, 2010.

[109]    Liu Y, Wang G, Zhu H Y, Ye Z Y. Numerical analysis of transonic buffet flow around a hammerhead payload fairing. Aerospace Science and Technology, 2019, 84: 604-619.

[110]    McCormick D C. Shock/boundary-layer interaction control with vortex generators and passive cavity. AIAA Journal, 1993, 31(1), 91-96.

[111]    Ogawa H, Babinsky H, Pätzold M, et al. Shock-wave/boundary-layer interaction control using three-dimensional bumps for transonic wings. AIAA Journal, 2008, 46(6): 1442-1452.

[112]    Dandois J, Lepage A, Dor J, Molton P. Open and closed-loop control of transonic buffet on 3D turbulent wings using fluidic devices. Comptes Rendus Mecanique, 2014, 342, 425-436.

[113]    Gao C Q, Zhang W W, Kou J Q, Liu Y L, Ye Z Y. Active control of transonic buffet flow. Journal of Fluid Mechanics, 2017, 824, 312-351.

[114]    Tian Y, Gao S, Liu P, et al. Transonic buffet control research with two types of shock control bump based on RAE2822 airfoil. Chinese Journal of Aeronautics, 2017, 30(5):1681-1696.

[115]    Giannelis N F, Vio G A, Levinski O. A review of recent developments in the understanding of transonic shock buffet. Progress in Aerospace Science, 2017, 92: 39-84.

[116]    Quan J G, Zhang W W, Gao C Q. Characteristic analysis of lock-in for an elastically suspended airfoil in transonic buffet flow. Chinese Journal of Aeronautics, 2016, 29(1), 129-143.